\documentclass[epj]{svjour}

\usepackage{amssymb}
\usepackage{amsmath}
\usepackage{graphicx}
\usepackage[caption=false]{subfig}
\usepackage{hyphenat}

\let\textAA\AA
\def\AA{\text{\textAA}}

\begin{document}
\title{Structural Relaxation and Aging Scaling in the Coulomb and Bose Glass Models}
\author{Hiba Assi$^{1,2}$ \and Harshwardhan Chaturvedi$^{1}$ \and Michel Pleimling$^{1,3}$ \and Uwe C. T\"{a}uber$^{1}$}
\institute{$^{1}$Department of Physics \& Center for Soft Matter and Biological Physics (MC 0435), Robeson Hall, 850 West Campus Drive, Virginia Tech, Blacksburg, Virginia 24061, USA\\
$^{2}$Physics and Engineering Department, Washington and Lee University, Lexington, VA 24450, USA\\
$^{3}$Academy of Integrated Science, College of Science (MC 0405), North End Center, Suite 4300, 300 Turner Street NW, Virginia Tech, Blacksburg, Virginia 24061, USA\\
  \email{hiba.assi@vt.edu; tauber@vt.edu}}
\date{Received: 10 June 2016 / Revised version: \today}



\abstract
{
We employ Monte Carlo simulations to study the relaxation properties of the two-dimensional Coulomb glass in disordered semiconductors and the three-dimensional Bose glass in type-II superconductors in the presence of extended linear defects. We investigate the effects of adding non-zero random on-site energies from different distributions on the properties of the correlation-induced Coulomb gap in the density of states (DOS) and on the non-equilibrium aging kinetics highlighted by the density autocorrelation functions. We also probe the sensitivity of the system's equilibrium and non-equilibrium relaxation properties to instantaneous changes in the density of charge carriers in the Coulomb glass or flux lines in the Bose glass.
}

\maketitle

\section{Introduction}
\label{sec:introduction}
Disordered semiconductors near a metal-insulator transition have been the focus of experimental and theoretical investigations for the past two decades~\cite{Efros1984,Efros1985,Pollak2013}. Experimental analysis of the conductivity of a two-dimensional silicon sample revealed non-trivial non-equilibrium relaxation properties: Jaroszynski and Popovi\'{c} drove the system out of equilibrium by altering the gate voltage, which is equivalent to suddenly changing the charge carrier density. They monitored the conductivity relaxation and discovered that the relaxation rate of the sample depended on the waiting time~\cite{popovic2007-1,popovic2007-2} before performing any measurements; this served as evidence for physical aging in such systems and therefore as an indication of their complex relaxation dynamics.

Our understanding of experimental results and theoretical studies investigating such strongly correlated disordered materials stems from the Coulomb glass model that was introduced by Efros and Shklovskii~\cite{Efros1984,Efros1985,Efros1975}. This model consists of charge carriers localized at random positions and interacting via strong long-range repulsive forces. Charge carriers are exposed to pinning defect sites and they reach thermal equilibrium by rearranging themselves via variable-range hopping. The strong anticorrelations originating from the long-range repulsive interactions cause the density of states to vanish near the chemical potential, creating a soft ``Coulomb gap" which in turn affects the carrier transport and equivalently the system's conductivity. The correlation-induced soft Coulomb gap was confirmed by electron tunneling experiments in doped semiconductors~\cite{Massey1995,Butko2000}. In addition to the correlation-induced equilibrium properties, this model displays complex non-equilibrium relaxation dynamics that has motivated careful theoretical investigations. Monte Carlo simulations performed by Grempel \textit{et al.}~\cite{Grempel2004,Kolton2005} discovered that the autocorrelation function measurements for different waiting times can be collapsed onto a master curve for various temperatures; an evidence of aging effects and specifically dynamical scaling~\cite{Henkel2010}. This theoretical study motivated Ovadyahu to investigate the emergence of aging effects in indium-oxide films after applying a ``quench-cooling'' protocol where the sample's temperature is lowered and the system then relaxes at the new temperature for a certain waiting time~\cite{Ovadyahu2006}. The gate voltage is then abruptly changed exciting the sample and the conductance is henceforth measured to confirm that the system's relaxation displays aging effects. Similar experimental work has been carried out by Grenet and Delahaye to study the out-of-equilibrium properties and specifically aging in disordered insulators at low temperatures, focusing on insulating granular aluminum thin films. They concluded that the sample's relaxation indeed depends on the waiting times since the quench-cooling application~\cite{Grenet2010}.

Disordered type-II superconductors similarly require deeper understanding due to the many competing energy scales and thus rich thermodynamic phases and transport properties~\cite{Blatter1994}. The most desirable property of type-II superconductors is the zero dissipation in current flow, hence one has to prevent vortex motion due to externally applied currents to eliminate the resulting Ohmic resistance. Optimal pinning mechanisms are utilized in these materials to prevent flux flow in the presence of applied currents. Among the different pinning configurations used, columnar defects have experimentally proven to be more efficient than uncorrelated point-like disorder~\cite{Civale1991}. At low temperatures, magnetic flux lines become attached to the entire length of these linear pinning sites. This results in the emergence of the strongly pinned ``Bose glass" phase~\cite{Fisher-Weichman1989,Kwok1992,Nelson1992,Lyuksyutov1992,Nelson1993} due to the localization of flux lines to these defect sites. 

Physical aging has been observed in some experiments involving superconducting materials. Du \textit{et al.} observed that the voltage response to an externally applied current in a superconducting 2H-NbSe$_{2}$ sample depended on the pulse duration, and that was an evidence of the occurrence of physical aging in disordered superconducting materials~\cite{Du2007}. On the numerical front, Pleimling and T\"{a}uber employed an elastic line model and Monte Carlo simulations to study magnetic flux lines in type-II superconductors analyzing their non-equilibrium relaxation properties, starting from somewhat artificial initial conditions where straight flux lines were randomly placed in the sample~\cite{Pleimling2011,Pleimling2015}. The resulting complex aging features (with identical initial conditions and
parameter values) were subsequently confirmed in a very different microscopic 
representation of the non-equilibrium vortex kinetics through Langevin 
molecular dynamics~\cite{Dobramysl2013}. Thereafter, Refs.~\cite{Assi2015,Assi2016} utilized these Langevin molecular dynamics simulations to investigate the effects of flux line density quenches on the lines' transverse fluctuations and found that the aging scaling exponents for these transverse fluctuations depended on the choice of initial conditions~\cite{Dobramysl2013,Assi2015,Assi2016}. The structural rearrangements of these vortex lines could not be probed in the computationally accessible time scales.

The Bose glass phase in type-II superconductors and the Coulomb glass model in semiconductors share many similarities~\cite{Nelson1992,Nelson1993,Dai1995,Tauber1995}. At low temperatures where the magnetic vortex lines are straight and at weak magnetic fields ($B<B_{\phi}$, with $B_{\phi}$ being the matching field), the Bose glass can be mapped to the Coulomb glass by considering the magnetic flux lines localized to columnar defects as particles pinned to point defects in a two-dimensional plane~\cite{Nelson1993,Dai1995}. The interacting flux lines in the Bose glass ``hop" from their occupied defect sites to surrounding unoccupied sites. This leads to the formation of double-kinks or ``superkink" excitations which are tongue-like double-kinks, each extending from one pinning site to another with a comparable energy~\cite{Nelson1992,Nelson1993}. Therefore, flux lines are not required to hop to nearest neighboring defect sites, which is similar to the variable-range hopping feature in the Coulomb glass. The phonon-mediated distance-depen-dent tunneling term in the Coulomb glass is replaced with an elastic energy term in the Boltzmann factor reducing the probability for long-distance double kinks in the Bose glass. Due to the remarkable similarity between the Coulomb and Bose glasses and the resulting mapping, one expects both systems to display similar properties. Analogous to the Coulomb glass, the Bose glass is composed of long-range interacting ``particles" and spatial disorder, and hence this model's density of states was confirmed to similarly display a soft gap~\cite{Dai1995,Tauber1995}, and the non-equilibrium relaxation dynamics display rich aging and scaling properties~\cite{Shimer2010,Shimer2014}. Numerical and analytical studies by Somoza \textit{et al.} investigated the density of states of two-dimensional systems with logarithmic interactions at zero and finite temperatures~\cite{Somoza2015}. They confirmed that their results from the two approaches are in perfect agreement at zero temperature. Furthermore, they observed that at finite temperatures the density of states is highly dependent on the system's temperature and can be scaled with this characteristic energy scale.

We employ Monte Carlo simulations to investigate the effects of random on-site energies and carrier density quen-ches on the density of states properties and aging scaling behavior in the Coulomb and Bose glass models. Our paper is organized as follows: The next section provides a brief theoretical background on the Coulomb glass model in disordered semiconductors and the Bose glass model in type-II superconductors in the presence of extended linear defects. It also discusses the Coulomb gap that forms in the density of states near the chemical potential due to the interaction-induced correlations. Section \ref{sec:model} presents an overview of the model we employ and the Monte Carlo algorithm we utilize to carry out our numerical investigations. Furthermore, it describes the protocol we perform, specifies the material parameter values we use, and presents the quantities we measure to analyze the system's equilibrium properties and non-equilibrium and aging properties. Section~\ref{sec:energy-analysis-1} demonstrates the effects of adding non-zero random on-site energies on the equilibrium and non-equilibrium relaxation properties in the Coulomb and Bose glass models. Section~\ref{sec:quench-analysis-2} analyzes the effects of sudden changes in the density of charge carriers in the Coulomb glass model or magnetic flux lines in the Bose glass model on the density of states properties and the non-equilibrium relaxation dynamics. We finally summarize our work in the concluding section~\ref{sec:con}.

\section{Theoretical Description}
\label{sec:theory}

In this section, we describe the Coulomb glass model and the properties we analyze throughout our investigation. We thereafter move to adapt the Coulomb glass model to the Bose glass in type-II superconductors in the presence of extended linear defects.

\subsection{Coulomb Glass Model}
\label{sec:coulomb-glass-theory}
The Coulomb glass model was introduced by Efros and Shklovskii in 1974~\cite{Efros1975} to describe localized charge carriers exposed to pinning defect centers in amorphous or doped crystalline semiconductors~\cite{Efros1984,Efros1985,Efros1975}. Since this model assumes that the localization length of charge carriers $\xi$ is of the same order or smaller than the mean separation $a_{0}$ between acceptor and donor sites, the carriers are confined to the randomly distributed sites. Furthermore, the carriers rearrange themselves to minimize the total interaction energy and thus reach thermal equilibrium at low temperatures. The carrier distribution rearrangements occur by variable-range hopping which is influenced by the phonon-assisted quantum tunneling between the acceptor and donor sites and thermally-induced jumps over energy barriers.

The system's Hamiltonian is given by~\cite{Efros1984,Efros1985,Efros1975} 
\begin{equation}
  \label{eq:coulomb-hamiltonian}
  \begin{split}
  H = \sum_{i}n_{i}\varphi_{i} + \frac{e^{2}}{2\kappa} \sum_{i\neq j}\frac{(n_{i}-K)(n_{j}-K)}{|\textbf{R}_{i}-\textbf{R}_{j}|} \, ,
  \end{split}
\end{equation}
where $e$ is the electron's charge, $\kappa$ is the dielectric constant, and $\textbf{R}_{i}$ and $n_{i}$ respectively denote the position vector and occupancy of defect site $i$ ($i=1,...,N$ with $N$ being the total number of defect sites), and the sites are randomly distributed on a square lattice~\cite{Yu1999}. The site occupancy is restricted to $n_{i}=0$ or $1$ due to the strong intra-site repulsions. $\varphi_{i}$ represents the $i$th site's bare energy, and thus the first term in (\ref{eq:coulomb-hamiltonian}) quantifies random site energies. The second term captures the repulsive Coulomb interactions. A uniform relative charge density $ K = \sum_{i} n_{i}/N$ is introduced to maintain global charge neutrality, where $K$ is equal to the total carrier density per site or the system's filling fraction. When we consider cases with all on-site energies $\varphi_{i}=0$, the system's Hamiltonian (\ref{eq:coulomb-hamiltonian}) is characterized by particle-hole symmetry: $K=0.5+k$ and $K=0.5-k$ (where $0 \leq k \leq 0.5$ is any deviation from the half-filled system) are equivalent. 

The resulting energy difference for a hop between two sites $i$ and $j$ is
\begin{equation}
  \label{eq:coulomb-energy-diff}
  \begin{split}
  \Delta E_{ij} = \epsilon_{j} - \epsilon_{i} - \frac{e^{2}}{\kappa R_{ij}} \, , 
  \end{split}
\end{equation}
where $R_{ij} = |\textbf{R}_{i}-\textbf{R}_{j}|$ is the distance between the two sites, and the interacting site energies are given by 
\begin{equation}
  \label{eq:coulomb-site-energy}
  \begin{split}
  \epsilon_{i} = \varphi_{i} + \frac{e^{2}}{\kappa} \sum_{j \neq i} \frac{n_{j} - K}{R_{ij}} \, .
  \end{split}
\end{equation}
It is worth noting that if we replace the sites occupation numbers with the Ising spin variables $\sigma_{i}= 2n_{i}-1 = \pm 1$, the Coulomb glass model would map into a random-site, random-field antiferromagnetic Ising model with long-range exchange interactions~\cite{Davies1982}. 

\subsection{Bose Glass Adaptation}
\label{sec:bose-glass-theory}

The Coulomb glass model above can be adapted to describe the low-temperature properties of magnetic flux lines in type-II superconductors in the presence of strong correlated columnar disorder~\cite{Nelson1993,Dai1995,Tauber1995}. These magnetic vortices become localized at the material's extended linear defects in the low-temperature Bose glass phase and subsequently thermal transverse fluctuations are suppressed. Therefore, the three-dimensional system becomes essentially two-dimensional by considering the cross-section of the superconductor where a pinning site is the location of the intersection of the superconductor's ``slice'' transverse to the magnetic field direction and a columnar defect. The charge carriers' role in the Coulomb glass model is played by the magnetic flux lines pinned to columnar defects in the Bose glass model. 

The mutual repulsive Coulomb interaction between two occupied sites is now replaced with a modified Bessel function $V(r) = 2 \epsilon_{0} K_{0}(r/\lambda)$, which is essentially a long-range logarithmic potential screened on the scale of the London penetration length $\lambda$. The Bose glass effective Hamiltonian is then 
\begin{equation}
  \label{eq:log-hamiltonian}
  \begin{split}
  H = \sum_{i}n_{i}\varphi_{i} + \epsilon_{0} \sum_{i\neq j} (n_{i}-K)(n_{j}-K) K_{0} ( R_{ij}/\lambda ) \, ,
  \end{split}
\end{equation}
where $\epsilon_{0}= (\phi/4\pi \lambda)^{2}$ represents the system's energy scale with the magnetic flux quantum $\phi=hc/2e$. In the current study, we only consider a dilute low-magnetic field regime where all distances between sites are $R_{ij} \ll \lambda$ and thus $K_{0}(x) \sim -\ln x$. 

Therefore, the energy difference for hops between two sites $i$ and $j$ in the Bose glass model $\Delta E_{ij}$ is
\begin{equation}
  \label{eq:log-energy-diff}
  \begin{split}
  \Delta E_{ij} = \epsilon_{j} - \epsilon_{i} - 2 \epsilon_{0} K_{0} ( R_{ij}/\lambda ) \, , 
  \end{split}
\end{equation}
with the interacting site energies 
\begin{equation}
  \label{eq:log-site-energy}
  \begin{split}
  \epsilon_{i} = \varphi_{i} + 2 \epsilon_{0} \sum_{j \neq i} (n_{j} - K) K_{0} ( R_{ij}/\lambda ) \, , 
  \end{split}
\end{equation}
where (similar to the Coulomb glass model) $\varphi_{i}$ and $n_{i}$ respectively denote the $i$th site random on-site energy and occupation number, and $K$ is the total vortex density per site; $K=1$ corresponds to $B=B_{\phi}$ with $B_{\phi}$ being the matching field.  

\subsection{Coulomb Gap Properties}
\label{sec:coulomb-gap-theory}
Efros and Shklovskii argued that the strong repulsive Cou-lomb interactions between charge carriers cause the system's density of states $g(\epsilon)$ to display a soft Coulomb gap in equilibrium~\cite{Efros1984}. Efros and Shklovskii's mean-field arguments yield that the density of states vanishes at the chemical potential $\mu_{C}$, which separates the low-energy $(\epsilon<\mu_{C})$ filled states and the higher-energy $(\epsilon>\mu_{C})$ empty states. For repulsive interactions of the form $V(r) \sim r^{-\sigma}$, they predict that the density of states follows a power law 
\begin{equation}
  \label{eq:density-power}
  \begin{split}
  g(\epsilon) \sim |\epsilon -\mu_{C}|^{\gamma}
  \end{split}
\end{equation}
for $\sigma<d$~\cite{Efros1984,Efros1985,Efros1975,Davies1982,Grunewald1982,Xue1988,Grannan1993,Menashe2000,Menashe2001,Surer2009,Mobius1992}, with a positive gap exponent $\gamma$. Their mean-field calculations, specifically applicable for wide random-site energy distributions \cite{Efros1979}, predict that the gap exponent is $\gamma_{mf} = (d/\sigma)-1$~\cite{Efros1985,Tauber1995}. These mean-field arguments assume single-particle hops and that all sites are evenly distributed within the energy interval $\epsilon$, whereas in fact multi-particle hops play an important role in the system and sites often cluster~\cite{Davies1982}. Therefore, $\gamma_{mf}$ represents a lower bound for the exponent $\gamma$~\cite{Efros2011}, for which M\"{o}bius, Richter, and Drittler obtained deviations from the predicted mean-field results~\cite{Mobius1992}. 

The existence of the correlation-induced Coulomb gap in the density of states significantly impedes the charge carrier mobility~\cite{Efros1975} effectively lowering the sample's conductivity: The gap exponent $\gamma$ alters the temperature dependence of the conductivity and the system's variable-range hopping length. In $d$ dimensions, the conductivity scales as $\ln \sigma \sim -T^{-p}$ where $p(\gamma) =(\gamma+1)/(\gamma+d+1)$ in the thermally-activated transport regime at low temperatures $T$. For a constant DOS, $\gamma=0$ and thus one recovers the Mott variable-range hopping exponent $p=1/(d+1)=1/4$ in three dimensions and $p=1/3$ in two dimensions. However, in the presence of the Coulomb gap, the non-zero gap exponent yields a larger $p$: $p=1/(1+\sigma)=1/2$ in three and two dimensions for $\sigma=1$, confirming that the Coulomb gap effectively lowers the low-temperature conductivity. 

Mean-field arguments predict an exponential suppression of the density of states in the Bose glass phase, corresponding to an infinite gap exponent value. Therefore, the Coulomb gap in the Bose glass model strongly suppresses flux creep and impedes flux transport resulting in the very favorable effect of a reduced resistivity which scales as $\ln \rho \sim -J^{-p}/T$ (with the driving currents $J$) in the thermally assisted creep regime~\cite{Tauber1995}.

\section{Model and Algorithm Description}
\label{sec:model}
We employ Monte Carlo (MC) simulations to investigate the Coulomb glass model in disordered semiconductors and the Bose glass model in type-II superconductors with correlated columnar defects. This section specifies the MC algorithm we have utilized to study both models, the system parameters we considered, and the quantities we measured.

\subsection{Model Description}
\begin{figure}
    \centering
    \subfloat{\includegraphics[width=0.97\columnwidth]{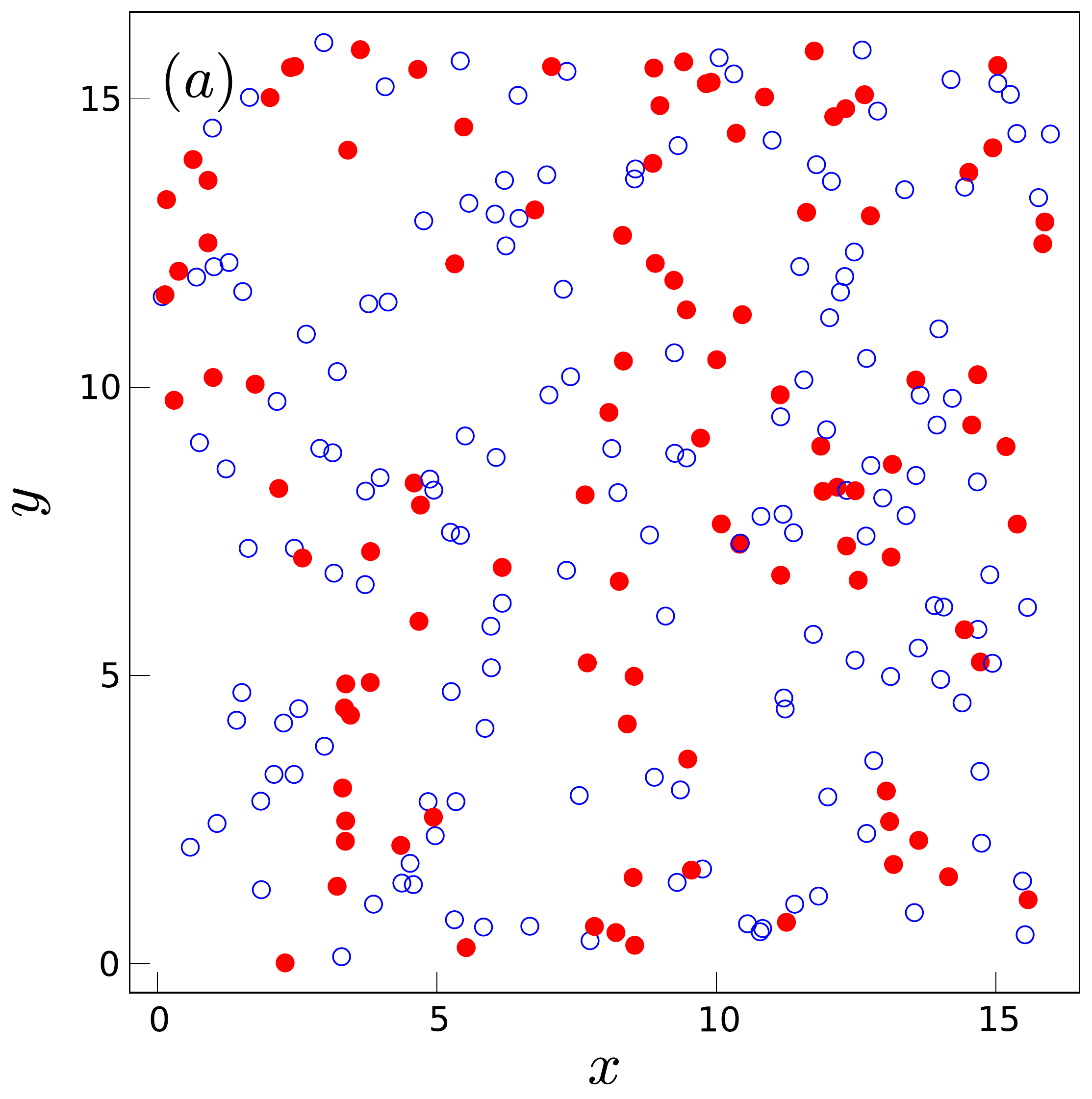} \label{fig:2d-coulomb-simulation-cell-t0}}\\[-0.5ex] 
    \subfloat{\includegraphics[width=0.97\columnwidth]{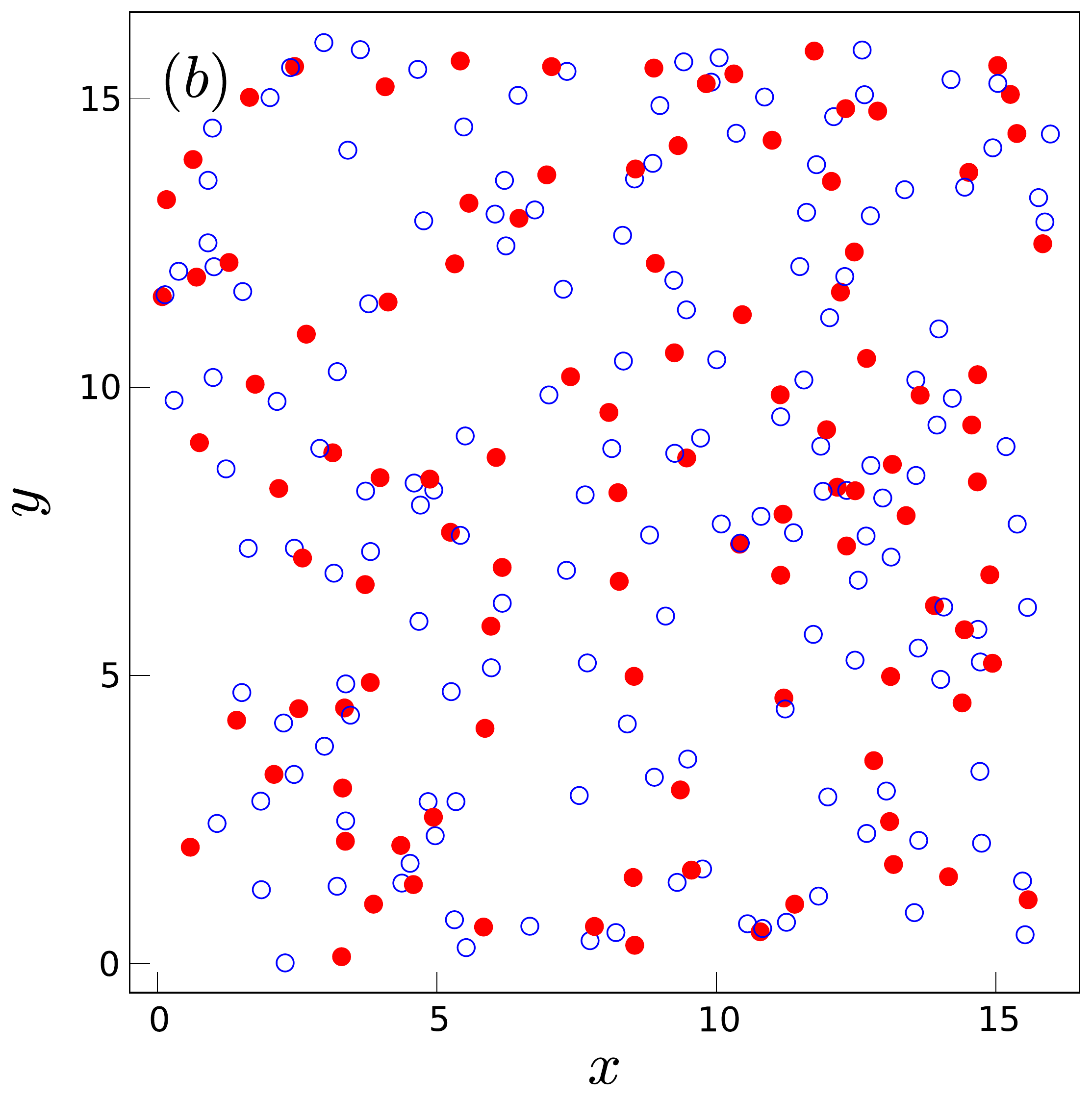} \label{fig:2d-coulomb-simulation-cell-t999}}
    \caption{An example layout in one of our simulations of the two-dimensional Coulomb glass with the filling fraction $K=0.5$ (a) initially and (b) after $10^{3}$ MC time steps when the system however is not yet fully relaxed. Open circles denote unoccupied pinning sites and filled circles denote occupied sites. (For illustration purposes, the point defects are given a finite width.)}
\label{fig:2d-coulomb-simulation-cell}
\end{figure}

Monte Carlo simulations are utilized to study the Coulomb and Bose glass models in $d=2$ dimensions. The simulation system is initiated by randomly placing $N$ pinning defect sites within a square lattice with periodic boundary conditions and charge carriers/flux lines that occupy $KN$ sites, where the site occupancy is restricted to $n_{i}=0$ or $1$. An example of the initial layout in one of our simulations is displayed in Fig.~\ref{fig:2d-coulomb-simulation-cell-t0}.

Then, charge carriers/flux lines attempt to hop from their corresponding occupied sites to unoccupied sites. The transition rate from site $i$ to site $j$ is determined by 
\begin{equation}
  \label{eq:hop-success}
  \begin{split}
  \Gamma_{ij}= \tau_{0}^{-1}e^{-2R_{ij}/\xi}\min[1,e^{-\Delta E_{ij}/k_{B}T}] \, ,
  \end{split}
\end{equation}
where $\tau_{0}$ denotes a microscopic time scale, $\xi$ represents the spatial extension of the charge carrier wave function or thermal wandering of the magnetic flux lines, and $\Delta E_{ij}$ given by Eq. (\ref{eq:coulomb-energy-diff}) for the Coulomb glass model or (\ref{eq:log-energy-diff}) for the Bose glass model is the difference between the energies of the two sites in question. Therefore, the success of hops is determined by two factors in (\ref{eq:hop-success}): an exponential term capturing the strongly distance-dependent phonon-mediated tunneling in semiconductors and vortex superkink proliferation in type-II superconductors, and a Metropolis term describing thermally-activated jumps through energy barriers. These hops allow the system in Fig.~\ref{fig:2d-coulomb-simulation-cell-t0} to relax to a new configuration shown in Fig.~\ref{fig:2d-coulomb-simulation-cell-t999} after $10^{3}$ MC time steps (the MC algorithm is outlined in the following section).

\subsection{Monte Carlo Algorithm}

One Monte Carlo time step (MCS) in our simulations occurs as follows:

\begin{enumerate}
\item Choose a random occupied defect site $i$.
\item Choose a random unoccupied defect site $j$ with probability $r_{1}=e^{-2R_{ij}/\xi}$.
\item Pick a random number $0 \leq r_{2} < 1$, and attempt a charge carrier hopping from $i$ to $j$. The probability of the transition success is given by the Metropolis term, $r_{3}=\min[1,e^{-\Delta E_{ij}/T}]$, \textit{i.e.}, move the charge carrier from defect site $i$ to site $j$ if $r_{2}<r_{3}$. If the hopping attempt fails, return to the first step.   
\end{enumerate}

One MCS consists of N iterations of the outlined steps. 

\subsection{Simulation Protocol}
The first system we study as a continuation to the work in Refs.~\cite{Shimer2010,Shimer2014} is the initial high-temperature quench with completely random initial conditions. The simulation lattice with the corresponding number of defect sites and carriers is prepared, and the system relaxes for a sufficiently long initial relaxation time for both models.

The second system we consider involves an abrupt change in the number of charge carriers, where carriers are randomly added or removed from the sample. This serves as an attempt to bridge our computational work with experimental research and to better understand the system's sensitivity to the choice of initial conditions. For this part, we let the system relax for a long time beyond any microscopic time scale $\tau_{0}$. At $t_{rel}=5 \cdot 10^{4}$ MCS, the number of charge carriers is instantaneously increased from $K=0.5$ to $K_{f}=0.54$ or lowered from $K=0.5$ to $K_{f}=0.46$. Then, we let the resulting system relax up to a waiting time $s$ measured after the quench. A snapshot of the system is taken at $s$ and the two-time carrier density correlation function defined in Section~\ref{sec:meas-quan} is measured at later times $t>s$. A similar study of flux line density quenches is performed to analyze structural rearrangements of flux lines in the Bose glass phase in disordered type-II superconductors.

\subsection{System Parameters}
All simulation distances are measured in units of the average distance between neighboring defect sites $a_{0}=1$. Simulation energies in the Coulomb glass model are measured in units of the typical Coulomb energy $E_{C}=e^{2}/\kappa a_{0}$, whereas energies in the Bose glass model are in units of $E_{B}=2\epsilon_{0}K_{0}(a_{0}/\lambda)$. Furthermore, all simulation times are measured in units of Monte Carlo time steps (MCS). The spatial extension of the charge carrier wave function is chosen to be $\xi=a_{0}$~\cite{Grempel2004,Kolton2005} following the careful investigation of different values of $\xi$ in Ref. \cite{Shimer2010}. In the Bose glass model, the London penetration length is set to $\lambda=8a_{0}$. 

Earlier work by Efros \textit{et al.}~\cite{Efros2011} focused on random site energies $\varphi_{i}$ from a uniform distribution of certain width, while here we investigate different energy distributions to conclude their effects on the system's properties. We consider cases with zero random on-site energies ($\varphi_{i}=0$) and others where the on-site energies are taken from either a normalized Gaussian or a flat distribution of zero mean and different widths $w$. 

All other simulation parameters are chosen in accordance with the findings in Ref. \cite{Shimer2010}. The simulation lattice length is $L=16$, since the study of various lengths proved the absence of measurable finite-size effects. The system's ambient temperature is set to $T=0.02$, since the temperature range that yields reasonable dynamics is $0.01<T<0.03$: the system's kinetics slowed down too much below $T=0.01$, and no aging was observed above $T=0.03$. The filling fraction is kept within the range $0.4<K<0.6$ since the dynamics in systems with filling fractions $K<0.4$ or $K>0.6$ freezes out in the accessible simulation time scales. Periodic boundary conditions are utilized, and each simulation configuration shown was analyzed with $1000$ to $6000$ independent simulation runs. 

\subsection{Measured Quantities}
\label{sec:meas-quan}
The density of states (DOS) $g(\epsilon)$, with the interacting site energies $\epsilon$ given by (\ref{eq:coulomb-site-energy}) for the Coulomb glass model and (\ref{eq:log-site-energy}) for the Bose glass model, is measured by finding the number of sites that are located in each energy bin, with the bin size chosen as $0.01$. We also study the soft gap that forms in the DOS and the speed of its formation over time given different simulation configurations. Mean-field calculations predict the DOS to display a power law near the chemical potential $\mu_{C}$, which directs us to measure the gap exponent $\gamma$ and compare it to the mean-field predicted $\gamma_{mf}$.
 
The non-equilibrium relaxation dynamics and response to the high-temperature quench and the charge carrier (or magnetic flux line) density abrupt changes are analyzed using the normalized two-time carrier density autocorrelation function
\begin{equation}
  \label{eq:cor-function}
  \begin{split}
  C(t,s) = \frac{\langle n_{i}(t)n_{i}(s)\rangle - K^2}{K(1-K)} \, ,
  \end{split}
\end{equation}
where $n_{i}(t)$ is the occupation number of defect site $i$ at time $t$, $s$ is the waiting time, and $K$ is the filling fraction. Furthermore, $\langle ... \rangle$ denotes an average over all sites, followed by an average over several thousand independent disorder realizations and thermal noise. 

\section{Random On-site Energy Effects}
\label{sec:energy-analysis-1}

Via Monte Carlo simulations, we have studied the emergence of the soft gap in the density of states in the Coulomb and Bose glass models in two dimensions. We compare the gap's properties and non-equilibrium relaxation dynamics of systems without random on-site energies to those with on-site energies from different distributions. Section~\ref{sec:energy-gap-analysis} discusses the effects of adding non-zero on-site energies on the density of states and the Coulomb gap's characteristics in the Coulomb and Bose glass models, while Section~\ref{sec:energy-aging-analysis} highlights the ensuing non-equilibrium relaxation dynamics and resulting aging scaling exponents. In this section, there are $N=256$ random pinning sites available to $KN=0.5 \times 256= 128$ charge carriers/flux lines.

\subsection{Coulomb Gap Properties}
\label{sec:energy-gap-analysis}

\subsubsection{Coulomb Glass Model}
\label{sec:gap-coulomb-analysis-1}

\begin{figure}[h]
  \centering
  \includegraphics[width=0.97\columnwidth]{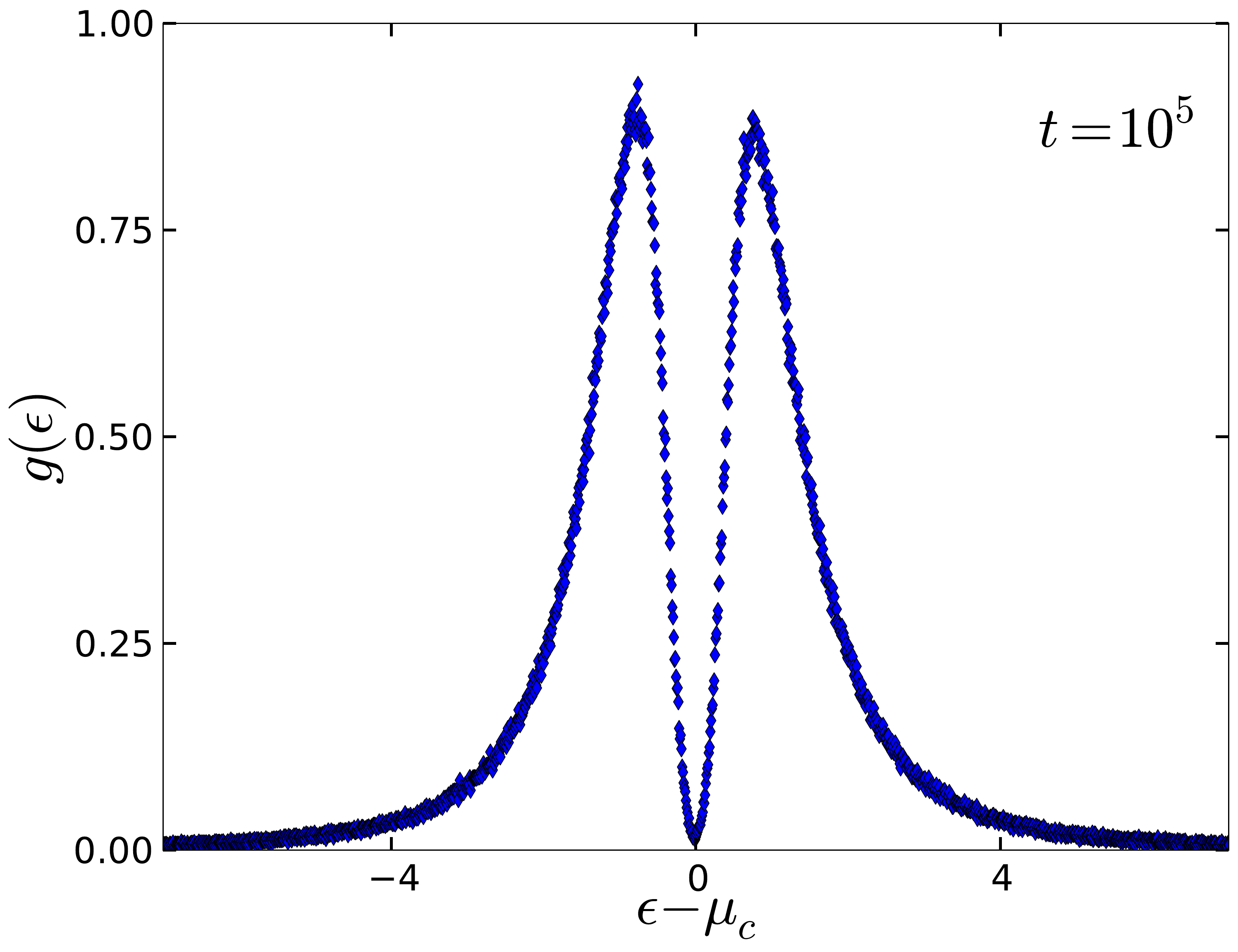}
  \caption{Density of states in the two-dimensional Coulomb glass model at $t=10^5$ MCS in the absence of random on-site energies. The density of states of this half-filled system is symmetric around the chemical potential (within statistical errors; data averaged over 6000 realizations).}
  \label{fig:coulomb-random-dos}
\end{figure}

We first investigate the Coulomb gap formation in the Coulomb glass model under random initial conditions in the absence of random on-site energies, where a fixed number of charge carriers is exposed to randomly distributed defect sites. At the beginning of our simulations before any hops occur, the density of states has a maximum at the chemical potential $\mu_{c}$. At $t=20$ MCS, a local minimum in the DOS already starts to form as a clear indication for the occupation of some low-energy states due to the charge carriers hopping between sites. The DOS in Fig.~\ref{fig:coulomb-random-dos} is almost totally suppressed at the chemical potential $\mu_{c}$ and the Coulomb gap is pronounced. This confirms the occupation of the low-energy states by the present charge carriers, which is pictured by the peak located at $\epsilon \leq \mu_{c}$, and the separation from the unoccupied higher-energy states at $\epsilon \geq \mu_{c}$.

\begin{figure}[h]
  \centering
  \includegraphics[width=0.97\columnwidth]{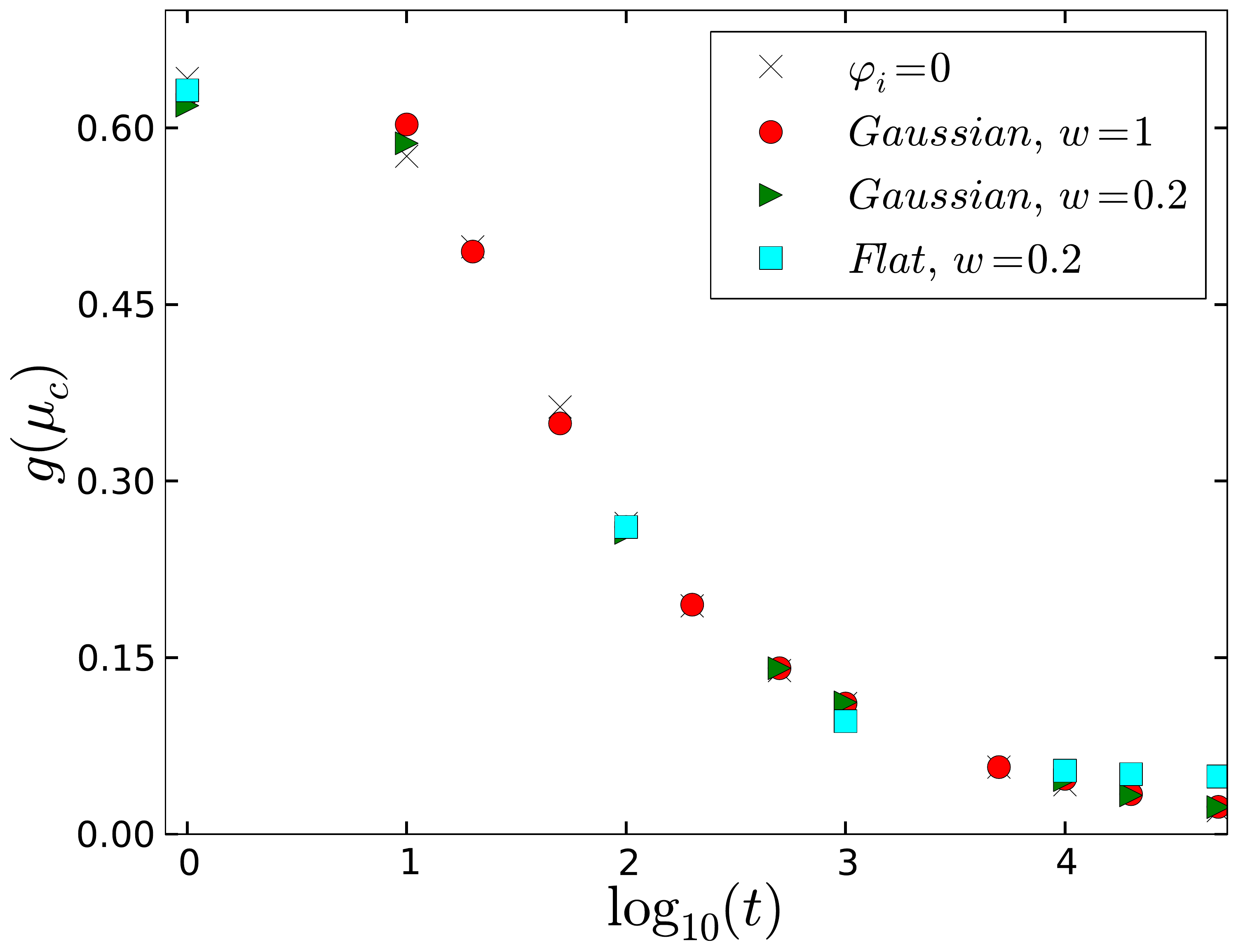}
  \caption{Coulomb gap formation in the two-dimensional Coulomb glass model in the presence of random on-site energies from different distributions (data averaged over 6000 realizations).}
  \label{fig:coulomb-random-onsite-density-evolution}
\end{figure}

We have monitored the evolution of the DOS in time in the absence of on-site energies to gauge the effect of the inclusion of such energies on the speed of the gap's formation. To this end, we studied systems without on-site energies and others with random on-site energies chosen from either a normalized Gaussian or a flat distribution of different widths. Fig.~\ref{fig:coulomb-random-onsite-density-evolution} shows the gap in the DOS starting to form within $t \sim 20$ MCS, and the value of $g(\mu_{c})$ decreasing as time progresses until (an almost) total suppression is observed in our two-dimensional Coulomb glass model. The three-dimensional system was observed in Refs. \cite{Shimer2010,Shimer2014} to display a faster total suppression of the Coulomb gap than the two-dimensional system.  

Fig.~\ref{fig:coulomb-random-onsite-density-evolution} confirms that the inclusion of random on-site energies from a Gaussian distribution of different widths (here $w=1$ and $w=0.2$ are utilized) does not influence the speed of the gap's formation and it does not affect the suppression of the DOS at the chemical potential at long times. On the other hand, one observes that on-site energies from a flat distribution of zero mean and width $w=0.2$ might be slowing down the suppression of the DOS at $\mu_{c}$, where $g(\mu_{c})$ reaches a plateau value after $10^{4}$ MCS of our accessible time scales as seen in Fig.~\ref{fig:coulomb-random-onsite-density-evolution}. It is possible that the inclusion of this energy distribution might have caused the density of states to freeze out in our small-sized Coulomb glass system for the accessible time scales considered here. 

\begin{figure}
  \centering
  \includegraphics[width=0.97\columnwidth]{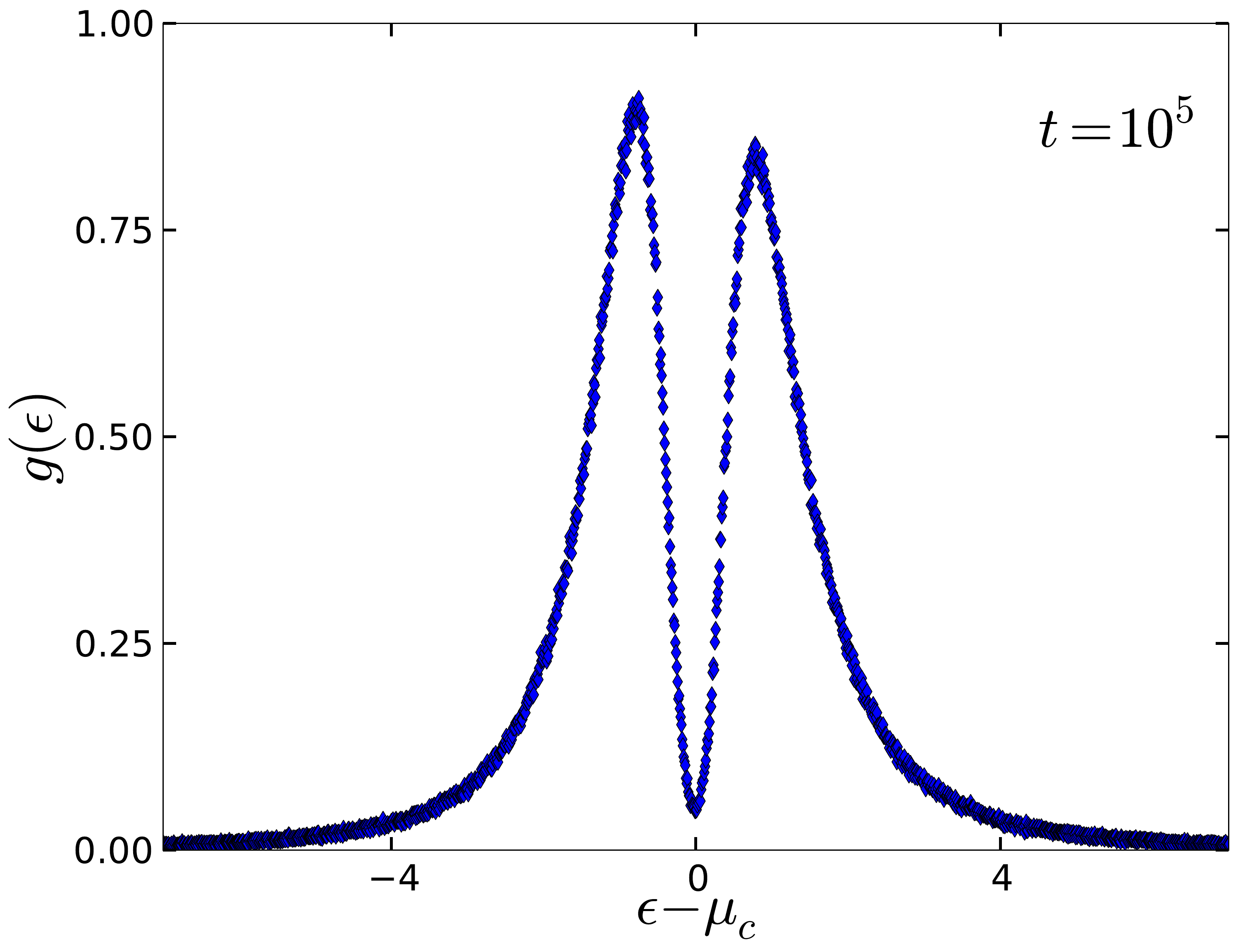}
  \caption{Density of states in the two-dimensional Coulomb glass model with random on-site energies from a flat distribution (zero mean, width $w=0.2$) at $t=10^5$ MCS (data averaged over 6000 realizations).}	
  \label{fig:coulomb-random-density-t100K-flat}
\end{figure}

Comparing Fig.~\ref{fig:coulomb-random-density-t100K-flat} to Fig.~\ref{fig:coulomb-random-dos} without on-site energies at the same time ($t=10^{5}$ MCS) asserts that on-site energies influence the arrangement of charge carriers and their pinning behavior to defect sites, thereby eliminating the symmetry of the DOS around $\epsilon=\mu_{c}$ for half-filling and more importantly delaying the time when total suppression of the DOS at $\epsilon=\mu_{c}$ is reached.

\begin{figure}
  \centering
  \includegraphics[width=0.97\columnwidth]{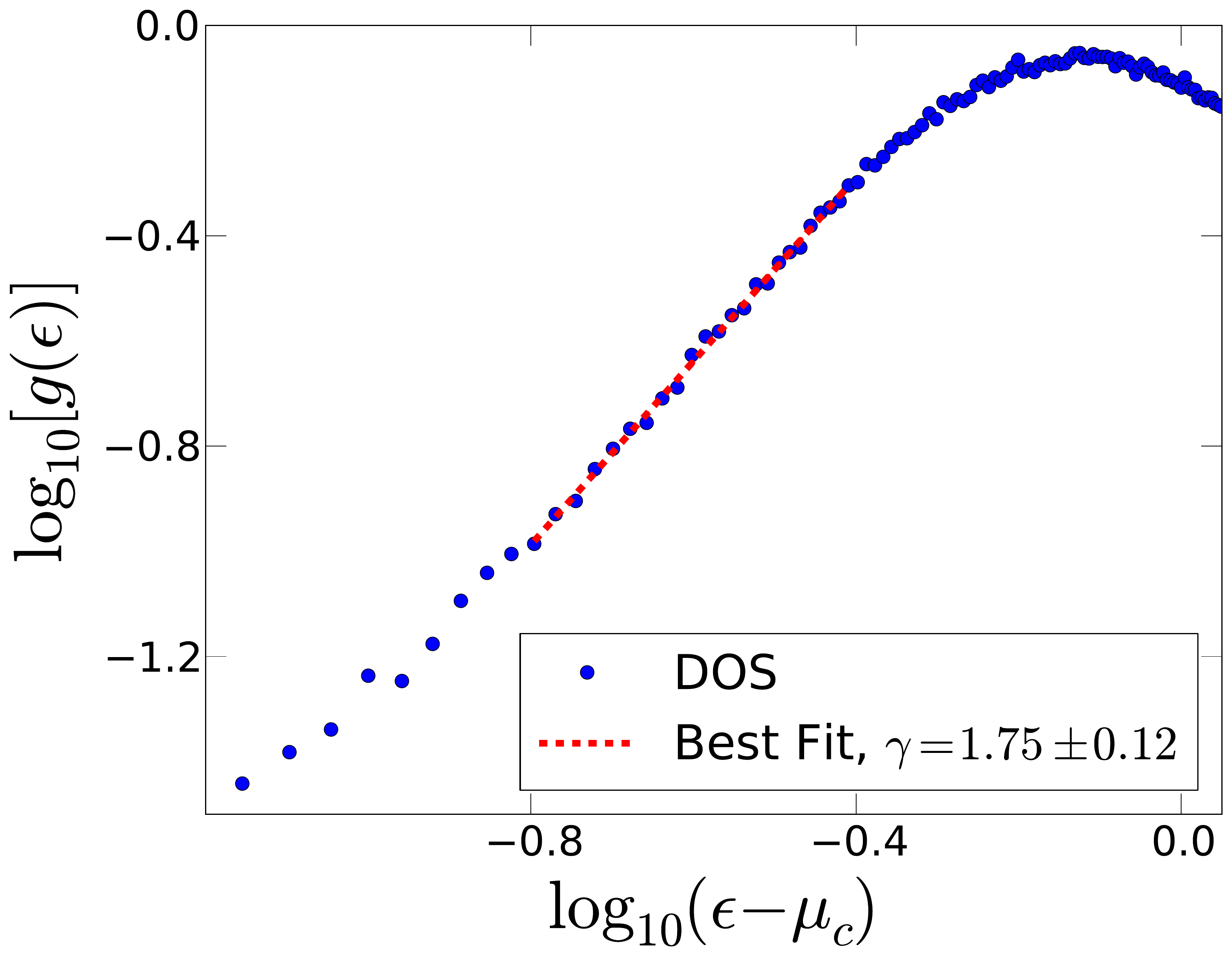}
  \caption{Logarithmic plot of the density of states in the two-dimensional Coulomb glass model at $t=10^{5}$ MCS in the absence of random on-site energies. The dashed line represents the best power law fit and the utilized data range from which it was extracted (data averaged over 6000 realizations).}	
  \label{fig:coulomb-random-exponent}
\end{figure}

To better understand the effect of non-zero random on-site energies on the Coulomb gap, we have studied the asymptotic behavior of the density of states near the chemical potential. As discussed earlier, the DOS near the chemical potential diminishes via a power law $g(\epsilon) \sim |\epsilon - \mu_{c}|^{\gamma}$ and mean-field arguments compute the gap exponent as $\gamma_{mf} = d-1$ for the Coulomb repulsive interactions of the form $1/r$. This implies that the mean-field gap exponent for the Coulomb glass model in two dimensions is $\gamma_{mf}=1$. We study the asymptotic behavior of the density of states near the chemical potential for the cases characterized by the absence and later presence of different on-site energies and observe that a power law behavior is displayed in the DOS near $\mu_{c}$; an example is shown in Fig.~\ref{fig:coulomb-random-exponent}.

As indicated in Fig.~\ref{fig:coulomb-random-exponent}, the computed effective gap exponent for the system without random on-site energies is found to be $\gamma=1.75$, which markedly differs from the mean-field prediction $\gamma_{mf}=1$. As discussed in Section~\ref{sec:theory}, it is expected for the exponent computed in these simulations to be different from that predicted by mean-field arguments. Furthermore, the mean-field prediction pertains to systems with strong disorder (wide random site energy distributions)~\cite{Efros1979}. We have measured the gap exponent in cases with different on-site energy distributions, and the values we found are summarized in Table~\ref{table:coulomb-random-exponent-table}.

Note that in order to estimate the characteristic error associated with the slope of the linear fit of our data, thus accounting for varying fluctuation sizes in different portions of the data, we divide the dataset into multiple regions. We first use a simple linear regression to obtain a line of best fit for the entire data range, producing the major line. We then divide the dataset into equal-sized regions along the x-axis and regressively fit a straight line to each region and note the slope of the line and standard deviation in the slope, two numbers that define an interval for the slope. This gives us two slope values associated with each region, \textit{i.e.}, the upper and lower bound of the interval. We then compare all the slope values obtained to the slope of the major line and report the largest difference as the characteristic error associated with the linear fit to our data.

The flat random on-site energy distribution has a much more pronounced influence on the value of the computed exponent than the Gaussian energy distributions. Including random on-site energies from this flat distribution decreases the gap exponent and renders it closer to the mean-field predictions, as expected~\cite{Efros1979}. Sharp energy distributions are less effective due to the interaction-induced correlations, and one obtains results that are very similar to the case when $\varphi_{i}=0$ (we confirmed this with flat and Gaussian distributions of zero mean and width $w=0.001$). In contrast, broader distributions with variances on the same scale as the correlation-induced Coulomb gap's width have a more pronounced effect on the density of states around the chemical potential. A lower effective gap exponent than the case without on-site energies implies a lower $p$, which in turn translates to an increased conductivity and thus charge carrier transport since conductivity scales as $\ln \sigma \sim -T^{-p}$.

\begin{table}[!htp]
\centering
\resizebox{\linewidth}{!}{%
\begin{tabular}{|c|c|c|c|c|}
\hline
 & (a)\,$\varphi_{i}=0$ &  (b)\,Gaussian &  (c)\,Gaussian &  (d)\,Flat\\
 & & $w=1$ & $w=0.2$ & $w=0.2$\\
\hline
$\gamma$ & $1.75\pm0.12$ & $1.62\pm0.18$ & $1.77\pm0.15$ & $1.26\pm0.13$\\ 
\hline
\end{tabular}}
\caption{Measured Coulomb gap exponent $\gamma$ for the two-dimensional Coulomb glass model (a) without random on-site energies, and in the presence of (b,c) Gaussian or (d) flat on-site energy distributions of different widths centered at zero. Exponents are computed from the best power law fit at $t=10^{5}$ MCS (data averaged over 6000 realizations).}
\label{table:coulomb-random-exponent-table}
\end{table}

\subsubsection{Bose Glass Model}
\label{sec:gap-bose-analysis-1}

\begin{figure}[h]
  \centering
  \includegraphics[width=0.97\columnwidth]{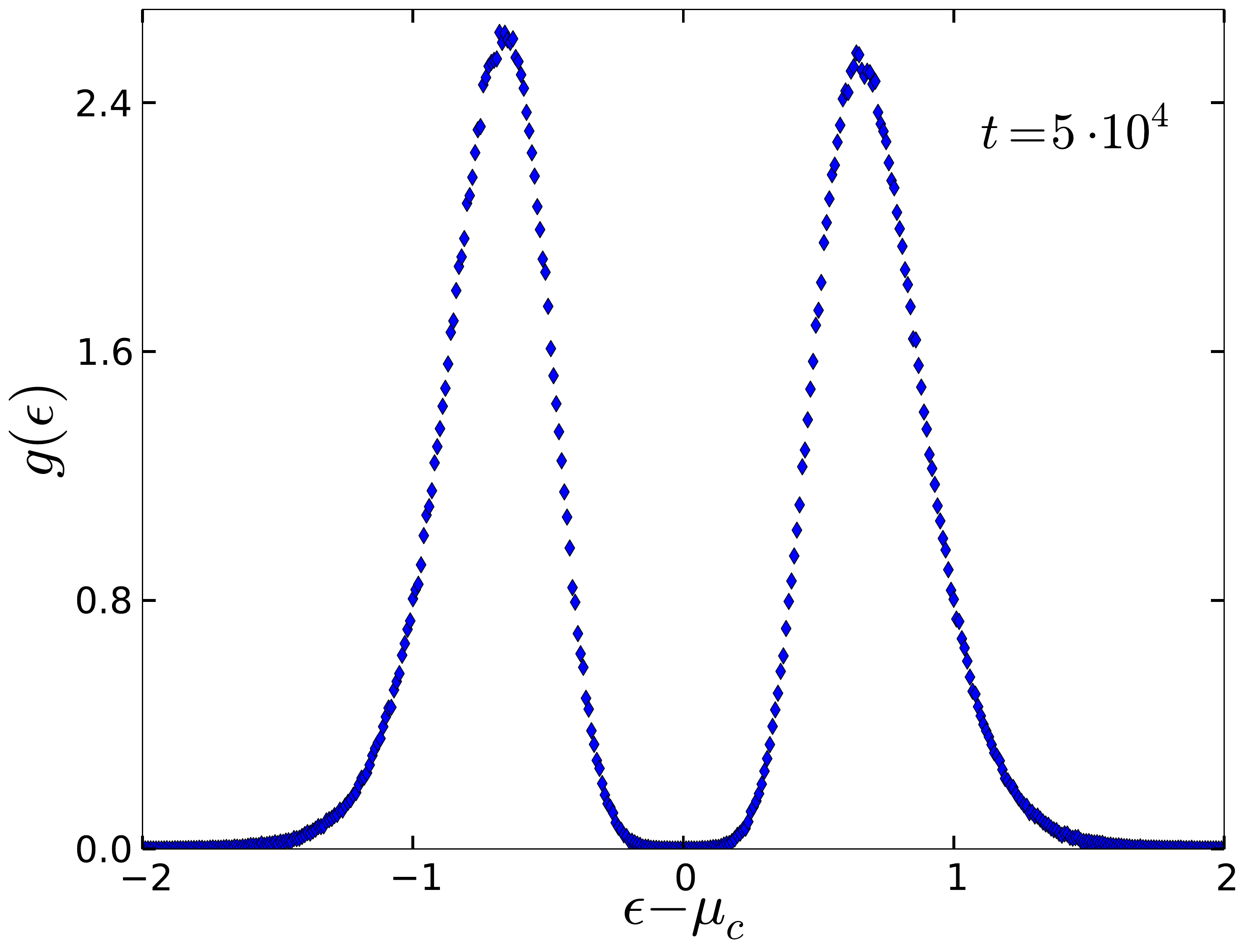}
  \caption{Density of states in the two-dimensional Bose glass model at $t=5 \cdot 10^4$ MCS in the absence of random on-site energies. The density of states of this half-filled system is symmetric around the chemical potential (within statistical errors; data averaged over 6000 realizations).}
  \label{fig:log-random-density-t50K}
\end{figure}

We have similarly monitored the density of states in the Bose glass model in two dimensions and analyzed the Coulomb gap's properties that characterize this system. A very pronounced difference with the Coulomb glass model is that the soft gap forms quicker and the DOS is suppressed faster, \textit{i.e.}, $g(\mu_{c})=0$ in shorter time, in about $50$ MCS. The total suppression of the DOS at $\mu_{c}$ for the Bose glass can be observed in Fig.~\ref{fig:log-random-density-t50K}. 

\begin{figure}[!ht]
  \centering
  \includegraphics[width=0.97\columnwidth]{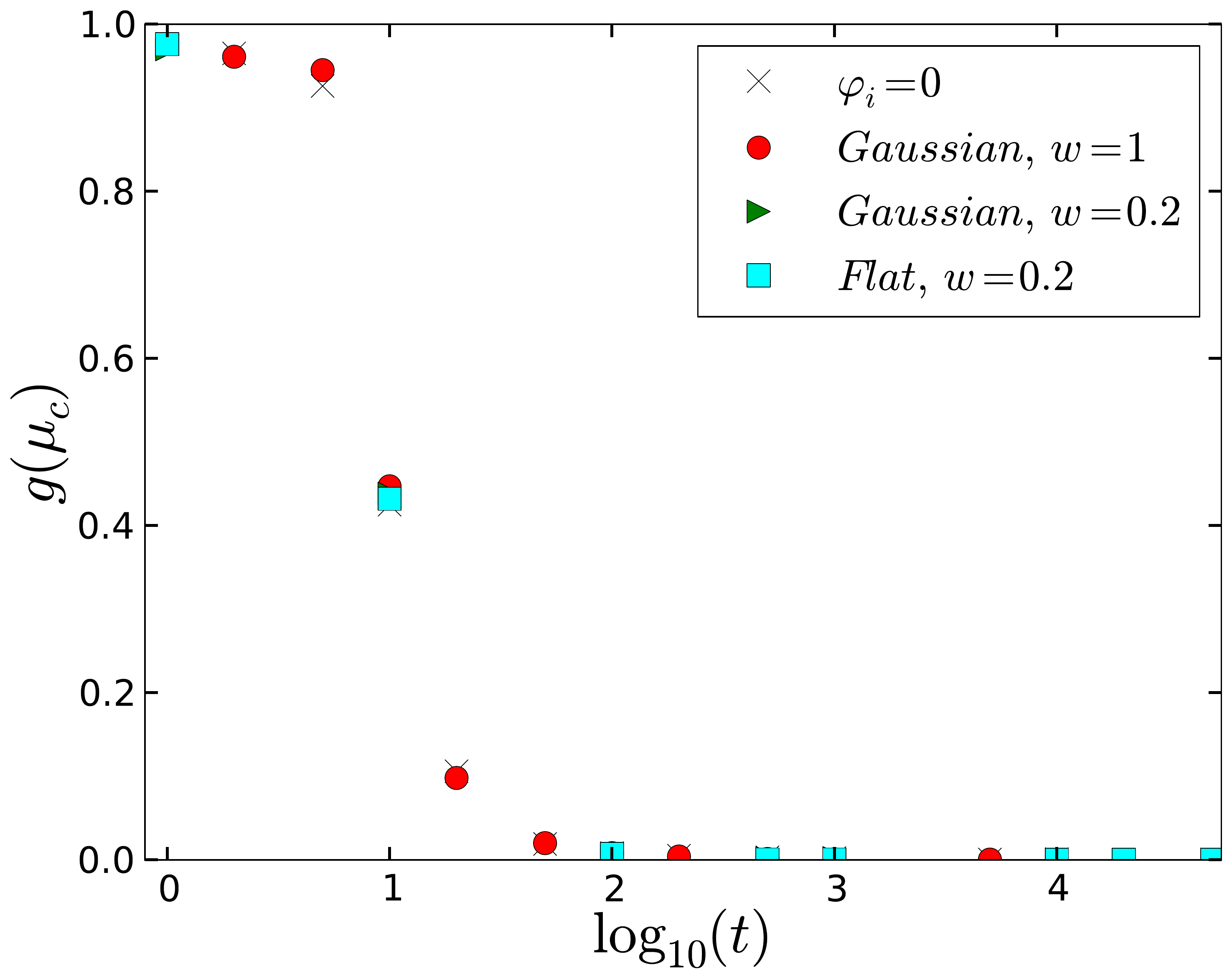}
  \caption{Coulomb gap formation in the two-dimensional Bose glass model in the presence of random on-site energies from different distributions (data averaged over 6000 realizations).}
  \label{fig:log-random-onsite-density-evolution}
\end{figure}

The Coulomb gap forms much faster and is broader than that displayed by systems with Coulomb repulsive interactions. Analyzing the evolution of $g(\mu_{c})$ in time confirms this observation and the direct effect of the addition of random on-site energies from various distributions. Comparing the speed of formation of the Coulomb gap in systems with logarithmic interaction displayed in Fig.~\ref{fig:log-random-onsite-density-evolution} to that in systems with Coulomb interactions in Fig.~\ref{fig:coulomb-random-onsite-density-evolution} proves that the logarithmic repulsive interactions drive the gap to form much faster.

Aside from the faster formation of the gap, the DOS is also strongly suppressed for systems without on-site energies and energies from flat and Gaussian distributions alike. Therefore, one concludes that the inclusion of on-site energies from different distributions does not change the overall properties and speed of formation of the gap in the Bose glass model, unlike our findings in the two-dimensional Coulomb glass model where energies from a flat distribution affect the suppression of $g(\mu_{c})$. To further investigate the effects of random on-site energies on the density of states in the Bose glass model, we have checked for the power law fit and the gap exponent in this case. Similar to the Coulomb glass model in Section~\ref{sec:gap-coulomb-analysis-1}, we observe that $g(\epsilon)$ follows a power law in the vicinity of the chemical potential $\mu_{c}$. 

However, the gap exponents computed in the Bose glass are large when compared to those in the Coulomb glass. One has to keep in mind that the suppressed motion within the Bose glass model leads to a slow relaxation and thus a large gap exponent. Calculating the gap exponent from the power law best fit allows us to study the influence of random on-site energies on the density of states. We have computed the Coulomb gap exponents from the power law best fit in the Bose glass system in the absence of any on-site energies and in the presence of random energies from a flat distribution of width $w=0.2$ and a Gaussian distribution of widths $w=0.2$ or $w=1$. The summary of our findings is presented in Table~\ref{table:log-random-exponent-table}.

\begin{table}[h]
\centering
\resizebox{\linewidth}{!}{%
\begin{tabular}{|c|c|c|c|c|}
\hline
 & (a)\,$\varphi_{i}=0$ &  (b)\,Gaussian &  (c)\,Gaussian &  (d)\,Flat\\
 & & $w=1$ & $w=0.2$ & $w=0.2$\\
\hline
$\gamma$ & $4.46\pm0.39$ & $4.53\pm0.35$ & $4.61\pm0.39$ & $5.26\pm0.41$\\ 
\hline
\end{tabular}}
\caption{Measured Coulomb gap exponent $\gamma$ for the two-dimensional Bose glass models (a) without random on-site energies, and in the presence of (b,c) Gaussian or (d) flat on-site energy distributions of different widths centered at zero. Exponents are computed from the best power law fit at $t=5\cdot10^{4}$ MCS (data averaged over 6000 realizations).}
\label{table:log-random-exponent-table}
\end{table}

On-site energies from a Gaussian distribution of different widths show a much weaker effect on the gap exponent than the flat distribution, similar to our findings in the Coulomb glass model. The introduction of random on-site energies from the utilized flat distribution increases the effective gap exponent rendering it closer to the infinite value that mean-field arguments predict in the Bose glass phase.

\subsection{Non-Equilibrium Relaxation Dynamics}
\label{sec:energy-aging-analysis}
In addition to our investigations of the effects of random on-site energies on the density of states, the Coulomb gap formation, and the effective gap exponent, we have analyzed the corresponding effects on the non-equilibrium relaxation properties of the systems studied following a quench from a completely uncorrelated, high-temperature initial state. Section~\ref{sec:aging-coulomb-analysis} presents our findings for the Coulomb glass model in disordered semiconductors, and Section~\ref{sec:aging-bose-analysis} summarizes our results for the Bose glass model in type-II superconductors in the presence of extended linear defects. 

\subsubsection{Coulomb Glass Model}
\label{sec:aging-coulomb-analysis}
We have investigated the non-equilibrium relaxation dynamics in the Coulomb glass model by measuring the two-time carrier density autocorrelation function $C(t,s)$ and its properties. 

\begin{figure}[!ht]
  \centering
  \includegraphics[width=0.97\columnwidth]{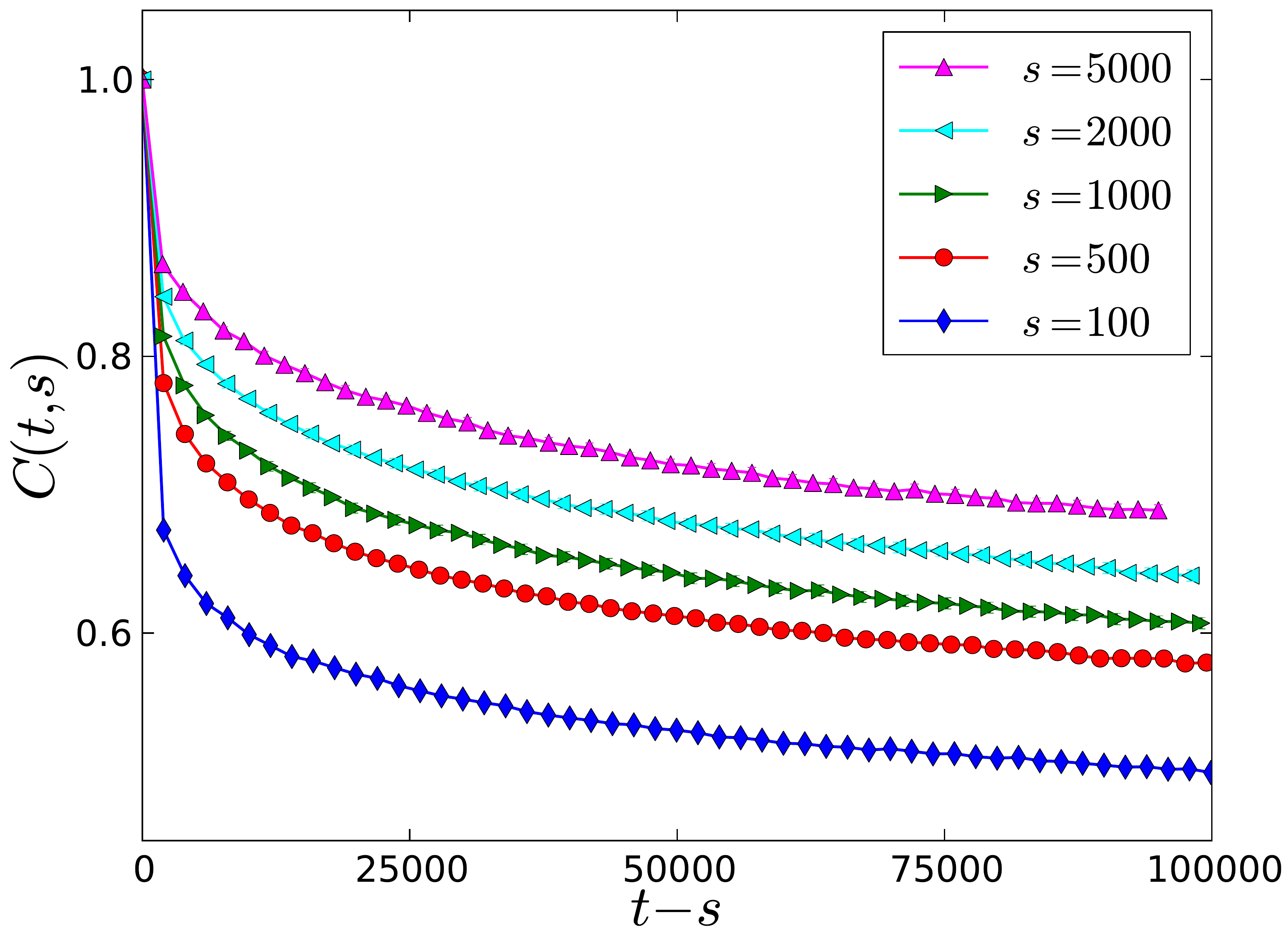}
  \caption{Relaxation of the two-time carrier density autocorrelation function in the two-dimensional Coulomb glass model with the filling fraction $K=0.5$ in the absence of random on-site energies (data averaged over 3000 realizations).}
  \label{fig:coulomb-random-c-linear}
\end{figure}

\begin{figure}[!ht]
  \centering
  \includegraphics[width=0.97\columnwidth]{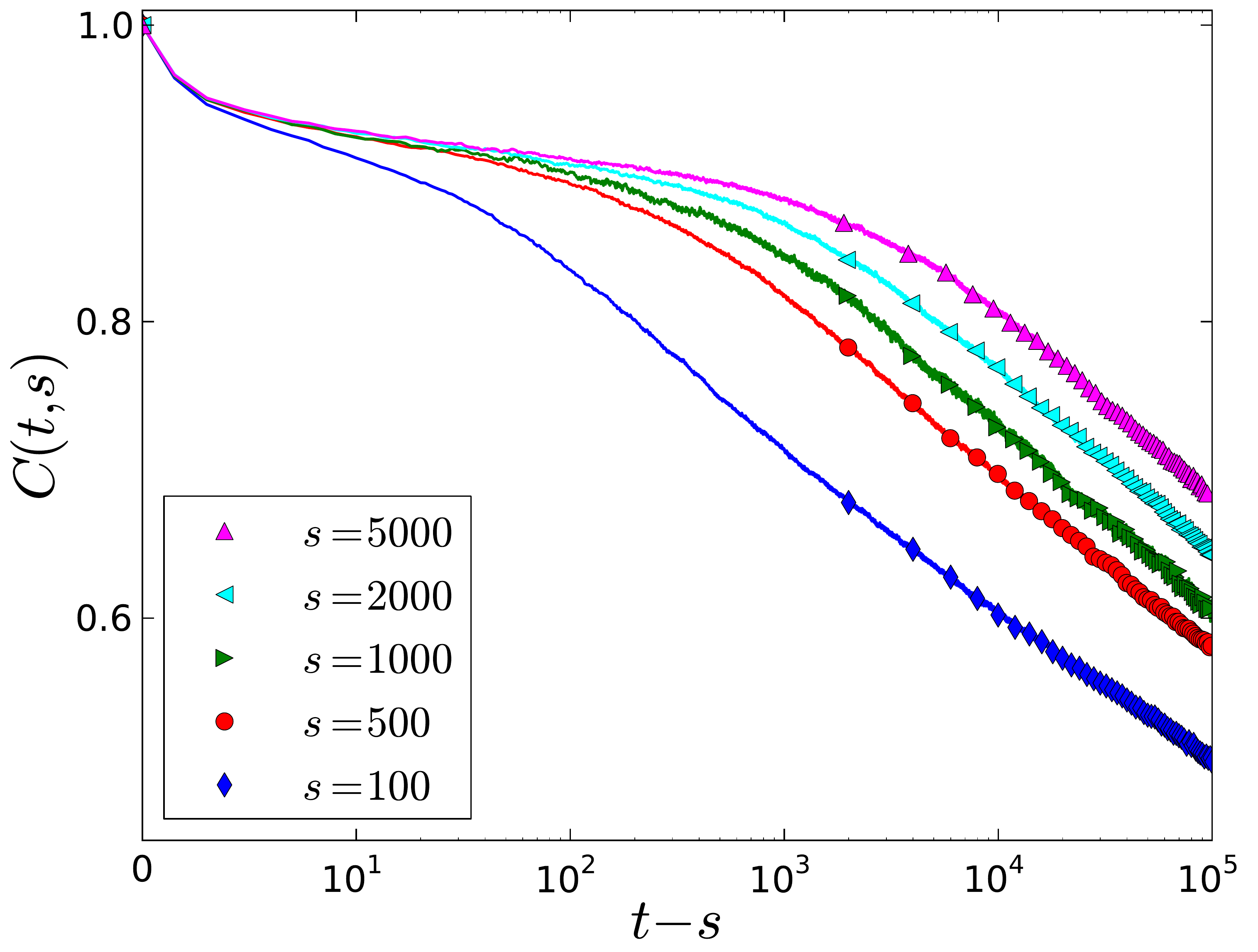}
  \caption{Relaxation of the two-time carrier density autocorrelation function in the two-dimensional Coulomb glass model with random on-site energies from a Gaussian distribution of zero mean and width $w=0.2$ (data averaged over 3000 realizations).}
  \label{fig:coulomb-gaus02-c-glassy}
\end{figure}

In the absence of random on-site energies, the system shows slow dynamics and breaking of time translation invariance due to its dependence on the waiting time$s$, as seen in Fig.~\ref{fig:coulomb-random-c-linear}. The slow relaxation of this system with Coulomb repulsive interactions is apparent in Fig.~\ref{fig:coulomb-random-c-linear} since the system has not reached a steady state even after $10^{5}$ MCS. The system displays two different relaxation regimes, similar to those shown in Fig.~\ref{fig:coulomb-gaus02-c-glassy} for the system with Gaussian on-site energies.

The first regime is characterized by a fast relaxation produced by the displacements of charge carriers around their initial positions and the carrier clusters' group motion. On the other hand, the second regime displays a slower relaxation that is dominated by the divisions of charge carrier clusters and the resulting relaxation of the system towards a steady state. The slow dynamics and breaking of time translation invariance are two characteristics of \textit{physical aging}, while the third property to be considered here is dynamical scaling~\cite{Henkel2010}. 

The carrier density autocorrelation function in the aging scaling regime obeys the general scaling form 

\begin{equation}
  \label{eq:C-full-aging-scaling-2}
  \begin{split}
C(t,s) = s^{-b}f_{C}(t/s) \, ,
\end{split}
\end{equation}
where $b$ is an aging scaling exponent. When $t/s \rightarrow \infty$ in the case of full aging, the scaling function $f_{C}$ is a power law $f_{C}(t/s) \sim (t/s)^{-\lambda_{C}/z}$, where $\lambda_{C}$ is the autocorrelation exponent and $z$ is the dynamic exponent~\cite{Henkel2010}.

\begin{figure}
    \centering
    \subfloat{\includegraphics[width=0.97\columnwidth]{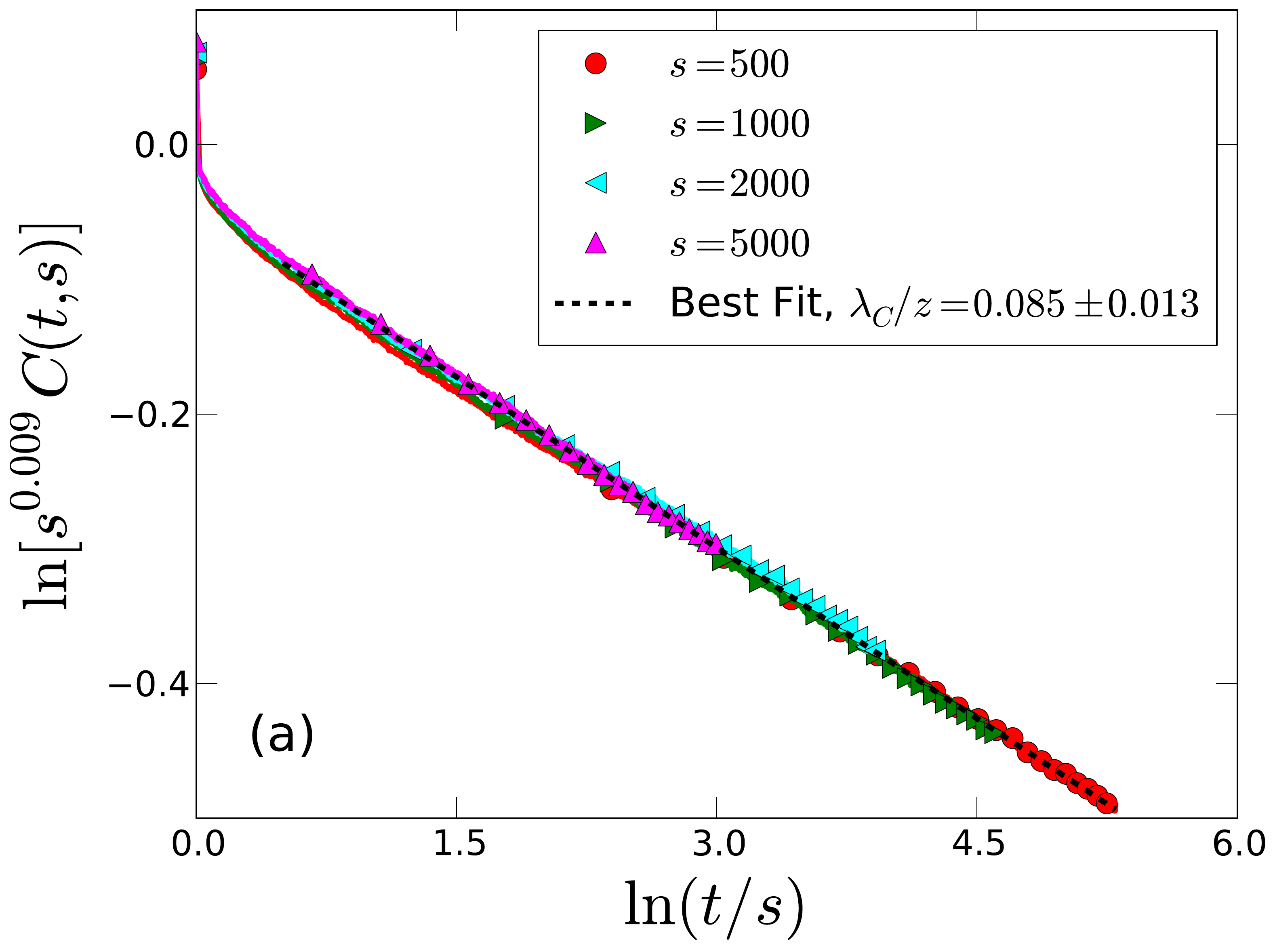} \label{fig:coulomb-random-c-aging}}\\[-0.5ex] 
    \subfloat{\includegraphics[width=0.97\columnwidth]{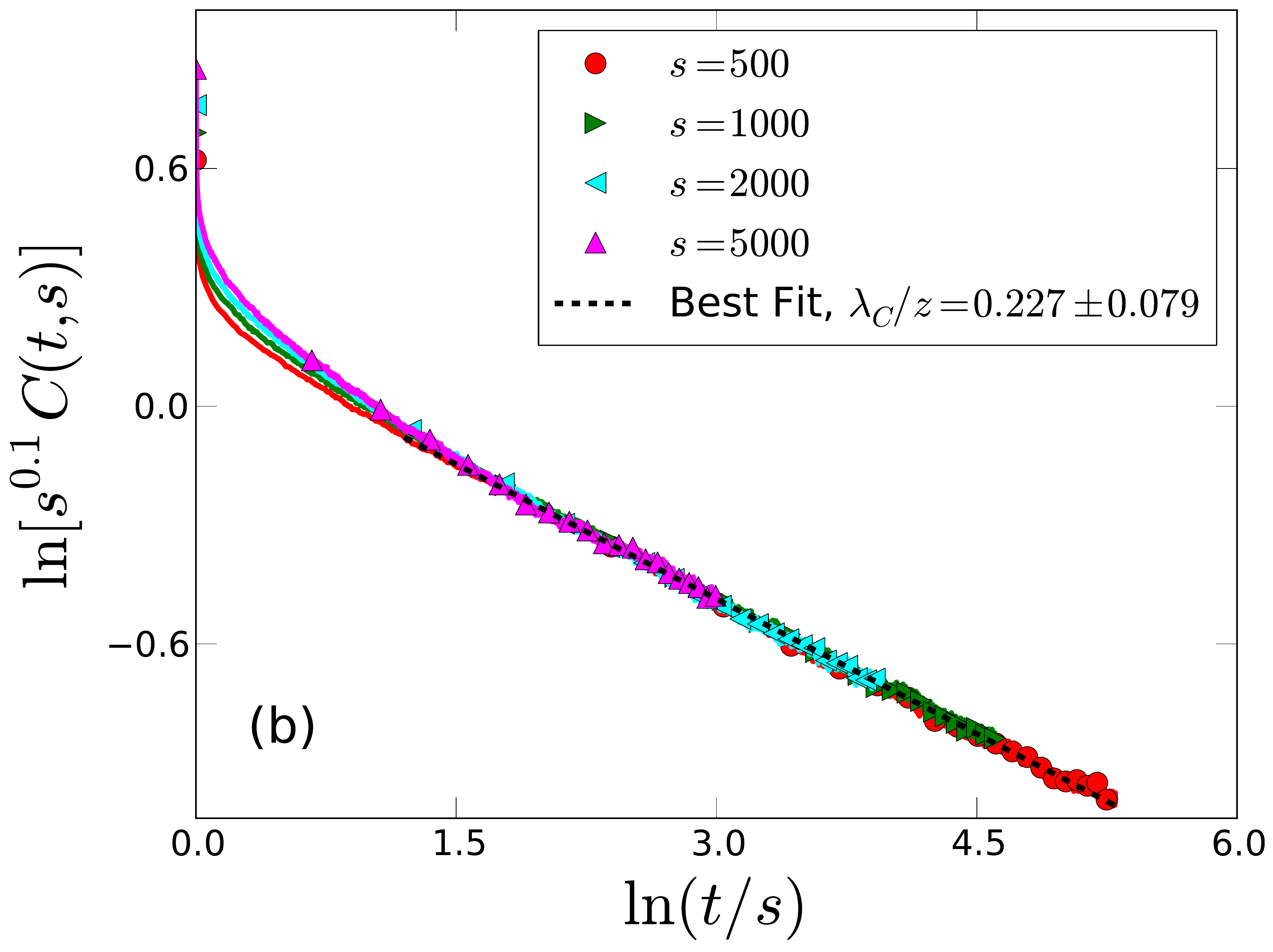} \label{fig:coulomb-flat02-c-aging}}
    \caption{Scaling of the two-time carrier density autocorrelation function in the two-dimensional Coulomb glass model (a) in the absence of random on-site energies and (b) with random on-site energies from a flat distribution of zero mean and width $w=0.2$ (data averaged over 3000 realizations).}
\label{fig:coulomb-c-aging}
\end{figure}

Investigating the dynamical scaling property in this regime, we observe that scaling collapse is obtained with the exponents $b=0.009$ and $\lambda_{C}/z=0.085$ for the two-dimensional Coulomb glass model without on-site energies, see Fig.~\ref{fig:coulomb-random-c-aging}. The goal of this section is to compute these scaling exponents after the addition of on-site energies from different distributions and compare them to the above case. We focus on energies from flat and Gaussian distributions, and we present the measured scaling exponents in Table~\ref{table:coulomb-random-scaling-exponent-table}.

\begin{table}[h]
\centering
\resizebox{\linewidth}{!}{%
\begin{tabular}{|c|c|c|c|c|}
\hline
 & (a)\,$\varphi_{i}=0$ &  (b)\,Gaussian &  (c)\,Gaussian &  (d)\,Flat\\
 & & $w=1$ & $w=0.2$ & $w=0.2$\\
\hline
$b$ & $0.009\pm0.002$ & $0.006\pm0.003$ & $0.01\pm0.002$ & $0.1\pm0.005$\\ 
\hline
$\lambda_{C}/z$ & $0.085\pm0.013$ & $0.082\pm0.018$ & $0.084\pm0.014$ & $0.227\pm0.079$\\
\hline
\end{tabular}}
\caption{Measured scaling exponents $b$ and $\lambda_{C}/z$ for the two-dimensional Coulomb glass model (a) without random on-site energies, and in the presence of (b,c) Gaussian or (d) flat on-site energy distributions of different widths centered at zero (data averaged over 3000 realizations).}
\label{table:coulomb-random-scaling-exponent-table}
\end{table}

From these results, we conclude that the addition of random on-site energies from a Gaussian distribution of different widths does not noticeably affect the aging scaling exponents. On the other hand, on-site energies from a flat distribution of width $w=0.2$ highly influence both exponents, see Fig.~\ref{fig:coulomb-flat02-c-aging}. The introduction of broad on-site energy distributions of widths on the same scale as the correlation-induced Coulomb gap's width drives the system to display a faster relaxation dynamics, as confirmed by the higher $\lambda_{C}/z$. 

\subsubsection{Bose Glass Model}
\label{sec:aging-bose-analysis}
A similar study of the non-equilibrium relaxation properties of the Bose glass model in two dimensions has been performed, where we have analyzed the effects of random on-site energies on the density autocorrelation function and its scaling form and exponents.

Starting with the Bose glass model with zero on-site energies, slow dynamics is also a signature of such a system and the autocorrelation function $C(t,s)$ displays a two-regime relaxation similar to that of the Coulomb glass model in Fig.~\ref{fig:coulomb-gaus02-c-glassy}. Investigating the universality of the scaling exponents, we study the full aging scaling forms for the various on-site energy distributions. 

To monitor the effects of on-site energies on the two-dimensional Bose glass system's non-equilibrium relaxation properties, we similarly investigate any discrepancies present between the scaling exponents in the case with zero on-site energies and those in the cases with different energy distributions. The results are summarized in Table~\ref{table:log-random-scaling-exponent-table}.

\begin{table}[h]
\centering
\resizebox{\linewidth}{!}{%
\begin{tabular}{|c|c|c|c|c|}
\hline
 & (a)\,$\varphi_{i}=0$ &  (b)\,Gaussian &  (c)\,Gaussian &  (d)\,Flat\\
 & & $w=1$ & $w=0.2$ & $w=0.2$\\
\hline
$b$ & $0.001\pm0.0008$ & $0.001\pm0.0008$ & $0.002\pm0.002$ & $0.04\pm0.004$\\ 
\hline
$\lambda_{C}/z$ & $0.057\pm0.009$ & $0.059\pm0.015$ & $0.059\pm0.017$ & $0.157\pm0.042$\\
\hline
\end{tabular}}
\caption{Measured scaling exponents $b$ and $\lambda_{C}/z$ for the two-dimensional Bose glass model (a) without random on-site energies, and in the presence of (b,c) Gaussian or (d) flat on-site energy distributions of different widths centered at zero (data averaged over 3000 realizations).}
\label{table:log-random-scaling-exponent-table}
\end{table}
\begin{figure}[!ht]
  \centering
  \includegraphics[width=0.97\columnwidth]{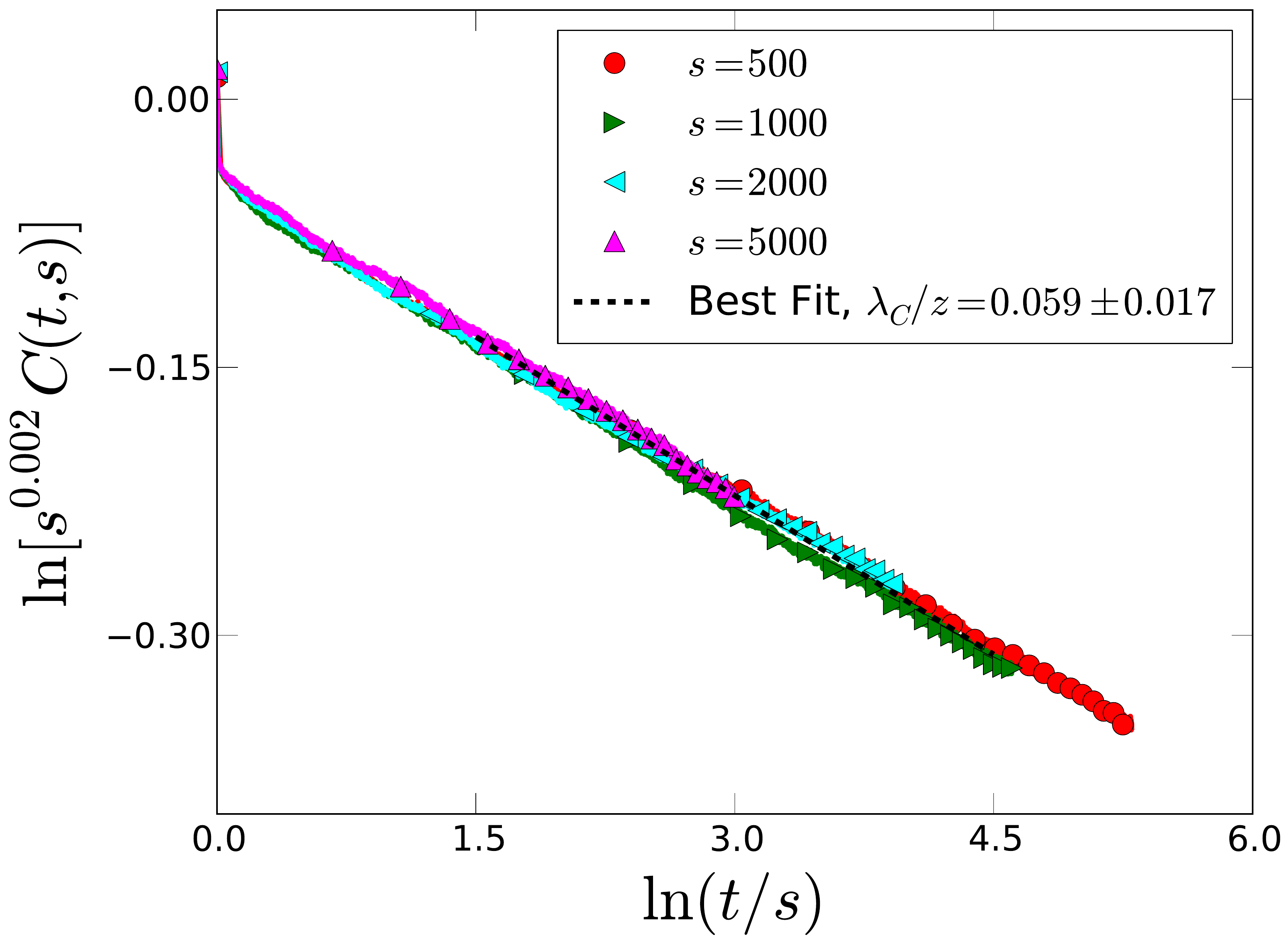}
  \caption{Scaling of the two-time density autocorrelation function in the two-dimensional Bose glass model with random on-site energies from a Gaussian distribution of zero mean and width $w=0.2$ (data averaged over 3000 realizations).}
  \label{fig:log-gaus02-c-aging}
\end{figure}
Our results in Table~\ref{table:log-random-scaling-exponent-table} confirm that adding random on-site energies from a Gaussian distribution of different widths has a minute  effect on the scaling exponents $b$ and $\lambda_{C}/z$, see Fig~\ref{fig:log-gaus02-c-aging}, which implies that this addition hardly modifies the relaxation dynamics of the Bose glass system with essentially logarithmic interactions. On the other hand, it is evident that a flat on-site energy distribution of width $w=0.2$ has a pronounced effect on the non-equilibrium relaxation properties of this system. Both aging scaling exponents $b$ and $\lambda_{C}/z$ assume considerably larger values as a response to the introduction of non-zero random on-site energies from a flat distribution, which hence imposes a faster relaxation on the system analogous to our results for the Coulomb glass model.

\section{Density Quench Effects}
\label{sec:quench-analysis-2}
Previous studies addressed the Coulomb gap and non-equilibrium relaxation properties under random initial conditions~\cite{Grempel2004,Kolton2005,Shimer2010,Shimer2014}. It is worthwhile to analyze the effect of abrupt changes in the density of charge carriers in the Coulomb glass model or analogously flux lines in type-II superconductors with linear defects on the Coulomb gap's properties and aging scaling exponents to probe the system's sensitivity to sudden changes in parameters. These quenches are experimentally achieved by quickly switching the gate voltage for semiconductors or the external magnetic field in type-II superconductors. We have utilized Monte Carlo simulations to study the effects of density quenches on the Coulomb gap in Section~\ref{sec:density-change-gap-analysis} and on the aging scaling behavior and exponents in Section~\ref{sec:density-change-scaling-analysis}, all in the absence of random on-site energies ($\varphi_{i}=0$).

\subsection{Coulomb Gap Properties}
\label{sec:density-change-gap-analysis}

\subsubsection{Coulomb Glass Model}
\label{sec:gap-coulomb-analysis-2}
In the Coulomb glass model in two dimensions, we investigate the effects of suddenly increasing the filling fraction from $K=0.5$ to $K_{f}=0.54$ (up-quench) or alternatively decreasing the filling fraction from $K=0.5$ to $K_{f}=0.46$ (down-quench) on the Coulomb gap and its evolution in time. 

We first keep the density of charge carriers fixed with the filling fraction $K=0.5$ and let the system relax in time for $t_{rel}=5 \cdot 10^{4}$ MCS until the Coulomb gap forms at the chemical potential $\mu_{c1}$ where $\mu_{c1}=0$ for half-filled systems (this is when the simulation clock is reset to $0$). After confirming that the density of states displays a relaxed Coulomb gap, we perform the density quench. Discussing first the case of the density up-quench and measuring the DOS, we have to keep in mind that the system is not at half-filling anymore but at $K_{f}=0.54$, which implies that we can expect a shift in the DOS as compared to the case with a fixed density. To test this expectation, we follow the DOS evolution in time. 

\begin{figure}
    \centering
    \subfloat{\includegraphics[width=0.97\columnwidth]{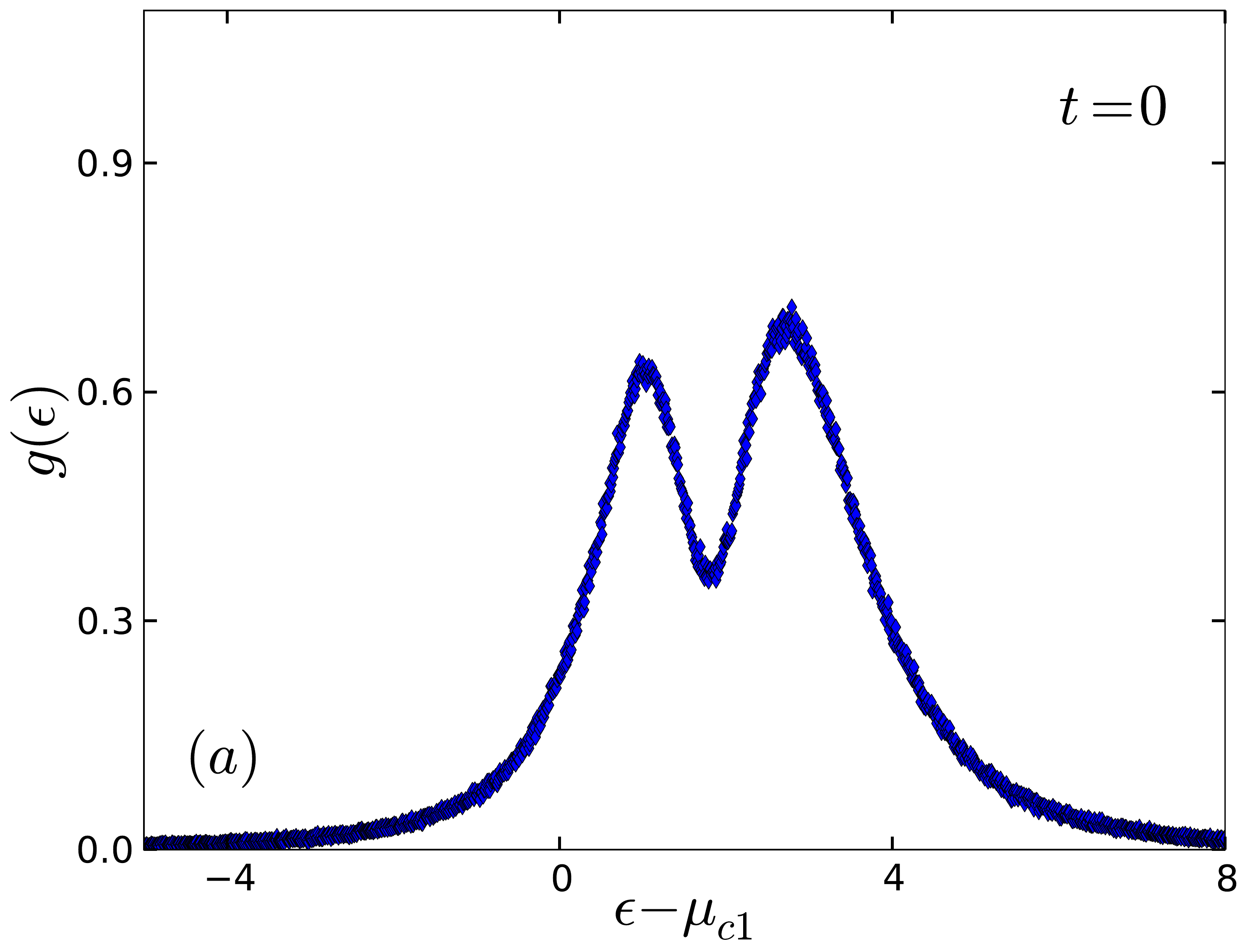} \label{fig:upquench-coulomb-gap-t0}}\\[-0.5ex] 
    \subfloat{\includegraphics[width=0.97\columnwidth]{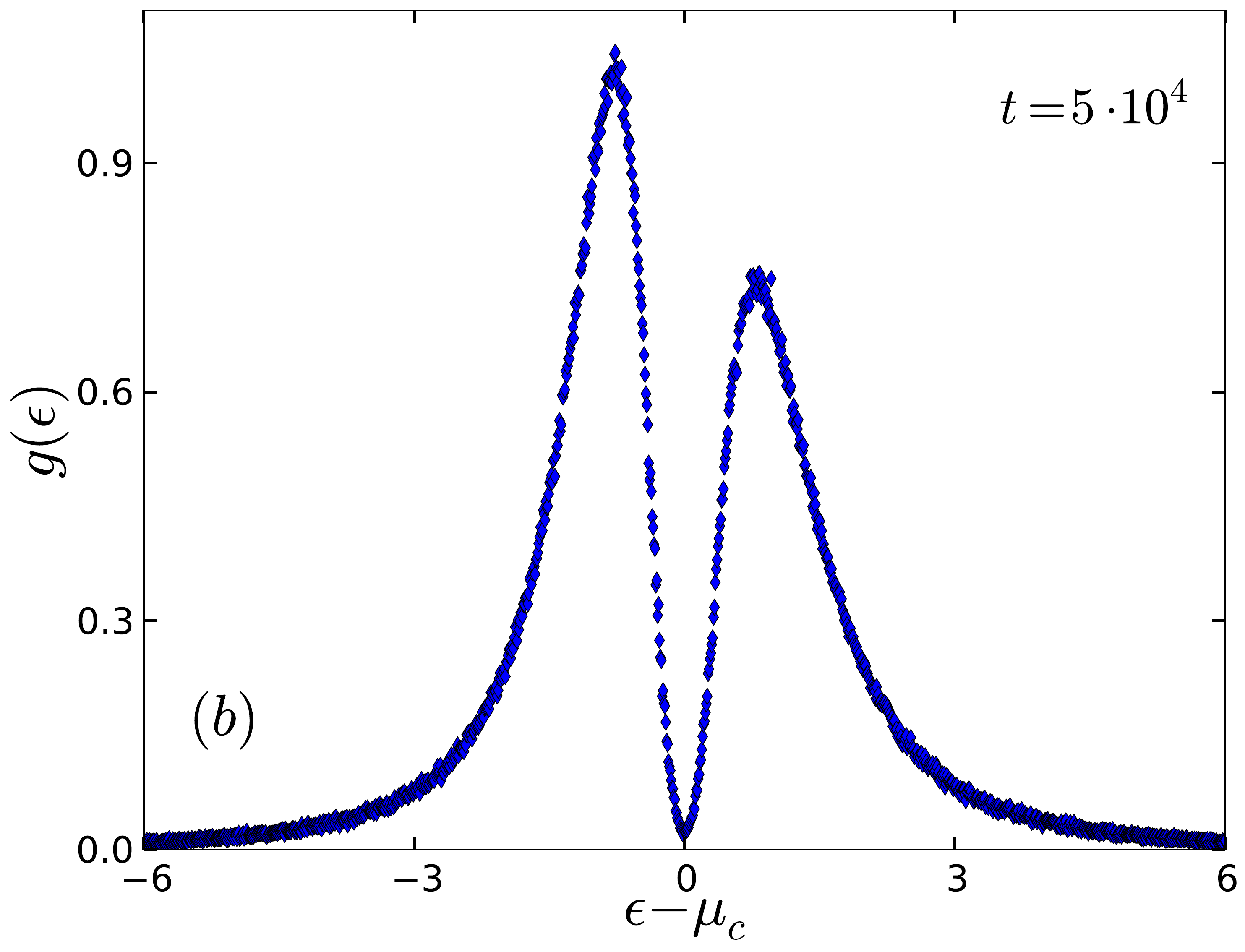} \label{fig:upquench-coulomb-gap-t50K}}
    \caption{Density of states in the two-dimensional Coulomb glass model with zero random on-site energies: (a) at $t=0$ at the moment of suddenly increasing the filling fraction from $K=0.5$ to $K_{f}=0.54$. $\mu_{c1}=0$ is the the relaxed system's chemical potential before the quench. (b) at $t=5 \cdot 10^{4}$ MCS after the sudden increase in the filling fraction. The equilibrated system's chemical potential after the quench is $\mu_{c}=2.11$ (data averaged over 6000 realizations).}
\label{fig:upquench-coulomb-gap}
\end{figure}

At the moment of the charge carrier density up-quench, Fig.~\ref{fig:upquench-coulomb-gap-t0} shows that the peak on the left denoting the lower-energy states reduces while the opposite peak increases. This occurs since new charge carriers are abruptly introduced to the system which in turn increases the typical Coulomb repulsion, and thus an enhanced peak on the higher-energy states initially emerges. As the system relaxes away from this initial arrangement and towards equilibrium, the new charge carriers start exploring their surroundings for energetically more favorable pinning sites. This results in a change of the shape of the DOS through a flip in the asymmetry, compare Fig.~\ref{fig:upquench-coulomb-gap-t0} with Fig.~\ref{fig:upquench-coulomb-gap-t50K}.

Since $K_{f} \neq 0.5$ after the density quench, this disappearance of the symmetry that characterizes the half-filled system is expected. In the case of an abrupt increase in $K$ away from half-filling, the equilibrium chemical potential is shifted to a new value $\mu_{c} > \mu_{c1}$. 

The case of density down-quench is similarly performed, where the filling fraction is abruptly lowered from $K=0.5$ to $K_{f}=0.46$. At the moment of the quench at $t=0$, the peak at the lower energies is enhanced, which is expected due to the reduction in the number of charge carriers which implies a decrease in the typical Coulomb repulsion. A similar but reversed asymmetry to that in Fig.~\ref{fig:upquench-coulomb-gap-t50K} therefore appears at $t=5\cdot10^{4}$ MCS after the down-quench. As the system relaxes, a smaller number of charge carriers occupies the original pinning defect sites, and hence the asymmetry is reversed to show an increased peak for the unoccupied states.   

Another property that can be utilized to investigate the effects of instantaneous density changes is the power law fit for the density of states in the vicinity of the chemical potential. We compute the effective gap exponent at the same time for the cases when the filling fraction is kept fixed at half or suddenly raised/lowered away from half-filling.

\begin{table}[h]
\centering
\resizebox{\linewidth}{!}{%
\begin{tabular}{llll}
\hline
\multicolumn{1}{|l|}{}         & \multicolumn{1}{l|}{(a)\,Fixed density} & \multicolumn{1}{l|}{(b)\,Up-quench} & \multicolumn{1}{l|}{(c)\,Down-quench} \\ \hline
\multicolumn{1}{|l|}{$\gamma$} & \multicolumn{1}{l|}{$1.75\pm0.12$}               & \multicolumn{1}{l|}{$1.68\pm0.12$}               & \multicolumn{1}{l|}{$1.68\pm0.14$}                 \\ \hline
                               &                                           &                                           &                                            
\end{tabular}}
\caption{Measured Coulomb gap exponent $\gamma$ for the two-dimensional Coulomb glass model with zero random on-site energies when the filling fraction is (a) kept fixed at $K=0.5$, (b) suddenly raised from $K=0.5$ to $K_{f}=0.54$, and (c) abruptly lowered from $K=0.5$ to $K_{f}=0.46$ at $t=5\cdot10^{4}$ MCS (data averaged over 6000 realizations).}
\label{table:updown-quench-coulomb-exponent-table}
\end{table}

From Table~\ref{table:updown-quench-coulomb-exponent-table}, one obtains the effective gap exponents for the three systems with different final filling fractions. It is worth noting that since the Coulomb gap exponent is an equilibrium property, one expects to obtain similar results when either starting from random initial conditions with filling fraction $K_{f}$ or abruptly changing the filling fraction from $K$ to $K_{f}$ and thereafter letting the system relax towards equilibrium.

\subsubsection{Bose Glass Model}
\label{sec:gap-Bose-analysis-2}
Having established that the DOS configuration is affected by sudden changes in the charge carrier density in the Coulomb glass model, we aim to carry out the same investigation for the Bose glass model with logarithmic interactions. This study is significant because it analyzes the effects of abrupt changes in the external magnetic field on the spatial rearrangements of magnetic vortex lines in type-II superconductors, while our earlier work in Ref.~\cite{Assi2015,Assi2016} focused on the flux lines height fluctuations and the structural relaxation time regime remained inaccessible. 

Similar to the study above, we start with the Bose glass model with a fixed filling fraction $K=0.5$, and we obtain the density of states in Fig.~\ref{fig:log-random-density-t50K}, where a Coulomb gap appears and the DOS is broad and totally suppressed at the chemical potential $\mu_{c}$, which implies that the system here is relaxed beyond microscopic time scales. In the case of suddenly increasing the system's filling fraction from $K=0.5$ to $K_{f}=0.54$, one notices the same behavior in the density of states in the Bose glass where the system starts with an enhanced peak at the higher-energy states, similar to Fig.~\ref{fig:upquench-coulomb-gap-t0}. Hence, the symmetry that is witnessed with a fixed number of magnetic flux lines at half-filling is broken due to the initial introduction of new magnetic flux lines into the system with a constant number of pinning centers. As the system relaxes to $t=5 \cdot 10^{4}$ MCS, the newly-added flux lines have explored the surrounding landscape and become pinned to the defect sites, and hence a peak on occupied sites is now enhanced. When we consider the consequences of abruptly reducing the number of flux lines due to changing the filling fraction from $K=0.5$ to $K_{f}=0.46$, we observe the same reversed asymmetry that is witnessed in the down-quench case in the Coulomb glass system.

\begin{table}[h]
\centering
\resizebox{\linewidth}{!}{%
\begin{tabular}{llll}
\hline
\multicolumn{1}{|l|}{}         & \multicolumn{1}{l|}{(a)\,Fixed density} & \multicolumn{1}{l|}{(b)\,Up-quench} & \multicolumn{1}{l|}{(c)\,Down-quench} \\ \hline
\multicolumn{1}{|l|}{$\gamma$} & \multicolumn{1}{l|}{$4.46\pm0.44$}               & \multicolumn{1}{l|}{$4.33\pm0.33$}               & \multicolumn{1}{l|}{$4.35\pm0.59$}                 \\ \hline
                               &                                           &                                           &                                            
\end{tabular}}
\caption{Measured Coulomb gap exponent $\gamma$ for the two-dimensional Bose glass model with zero random on-site energies when the filling fraction is (a) kept fixed at $K=0.5$, (b) suddenly raised from $K=0.5$ to $K_{f}=0.54$, and (c) abruptly lowered from $K=0.5$ to $K_{f}=0.46$ at $t=5 \cdot 10^{4}$ MCS (data averaged over 6000 realizations).}
\label{table:updown-quench-log-exponent-table}
\end{table}

We studied the effects sudden changes in the number of flux lines have on the effective gap exponent in the Bose glass model. The results are summarized in Table~\ref{table:updown-quench-log-exponent-table}.
The effective gap exponents for the three different final filling fractions in Table~\ref{table:updown-quench-log-exponent-table} are almost equal, which implies that this equilibrium property is similar in systems with the studied filling fractions $K=0.5$, $0.54$, and $0.46$ in the Bose glass model.

\subsection{Non-Equilibrium Relaxation Dynamics}
\label{sec:density-change-scaling-analysis}
In addition to studying the effects of charge carrier/flux line density quenches on the density of states and the soft Coulomb gap, we analyzed the effects of sudden addition/removal of carriers on the non-equilibrium relaxation properties and the aging scaling exponents: Section~\ref{sec:aging-coulomb-analysis-2} presents the results of this study in the Coulomb glass model and Section~\ref{sec:aging-Bose-analysis-2} summarizes our findings for the Bose glass model, all in the absence of random on-site energies ($\varphi_{i}=0$).

\subsubsection{Coulomb Glass Model}
\label{sec:aging-coulomb-analysis-2}
In the two-dimensional Coulomb glass model, we relax the system for $t_{rel}=5 \cdot 10^{4}$ MCS (this is when the simulation clock is reset to $0$) to then quench the density of charge carriers and measure the resulting two-time carrier density autocorrelation function for different waiting times $s$ measured after the quench. 
\begin{figure}[h]
  \centering
  \includegraphics[width=0.97\columnwidth]{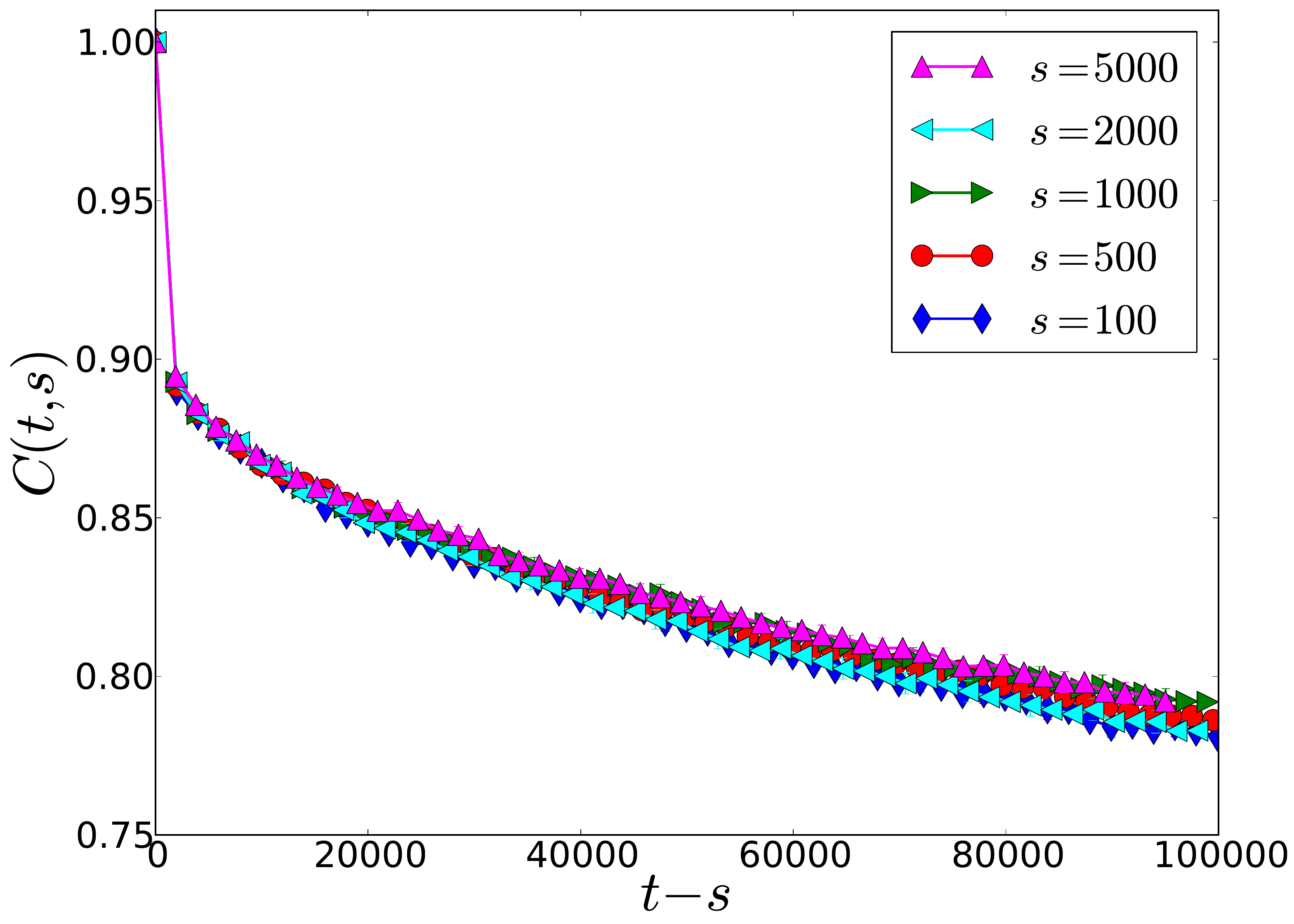}
  \caption{Relaxation of the two-time carrier density autocorrelation function in the two-dimensional Coulomb glass model with a fixed charge carrier density $K=0.5$ (data averaged over 2000 realizations).}
  \label{fig:noquench-coulomb-c-linear}
\end{figure}

In the case where the density is kept fixed at $K=0.5$, the dynamics in Fig.~\ref{fig:noquench-coulomb-c-linear} becomes less influenced by the choice of waiting times due to the long initial relaxation time that this finite system undergoes. The density autocorrelation function of this relaxed system with a fixed $K$ does not acquire the dynamical scaling property, regardless of the choice of scaling exponents, possibly due to the finite-size effects that emerge at the time scales considered. When the charge carrier density suddenly changes, we observe different features to the case with a fixed filling fraction $K$. 

\begin{figure}[h]
    \centering
    \subfloat{\includegraphics[width=0.97\columnwidth]{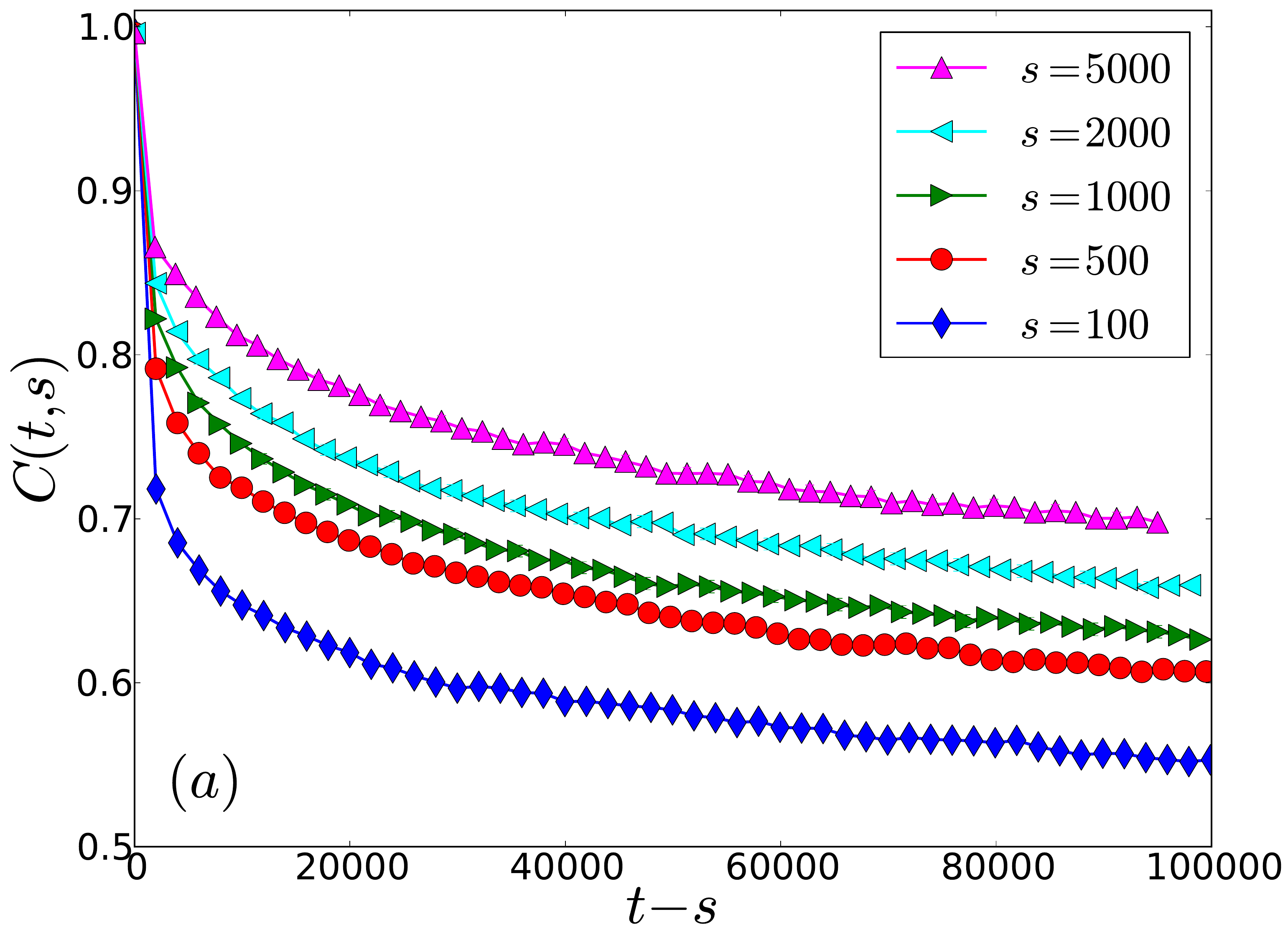} \label{fig:upquench-coulomb-c-linear}}\\[-0.5ex] 
    \subfloat{\includegraphics[width=0.97\columnwidth]{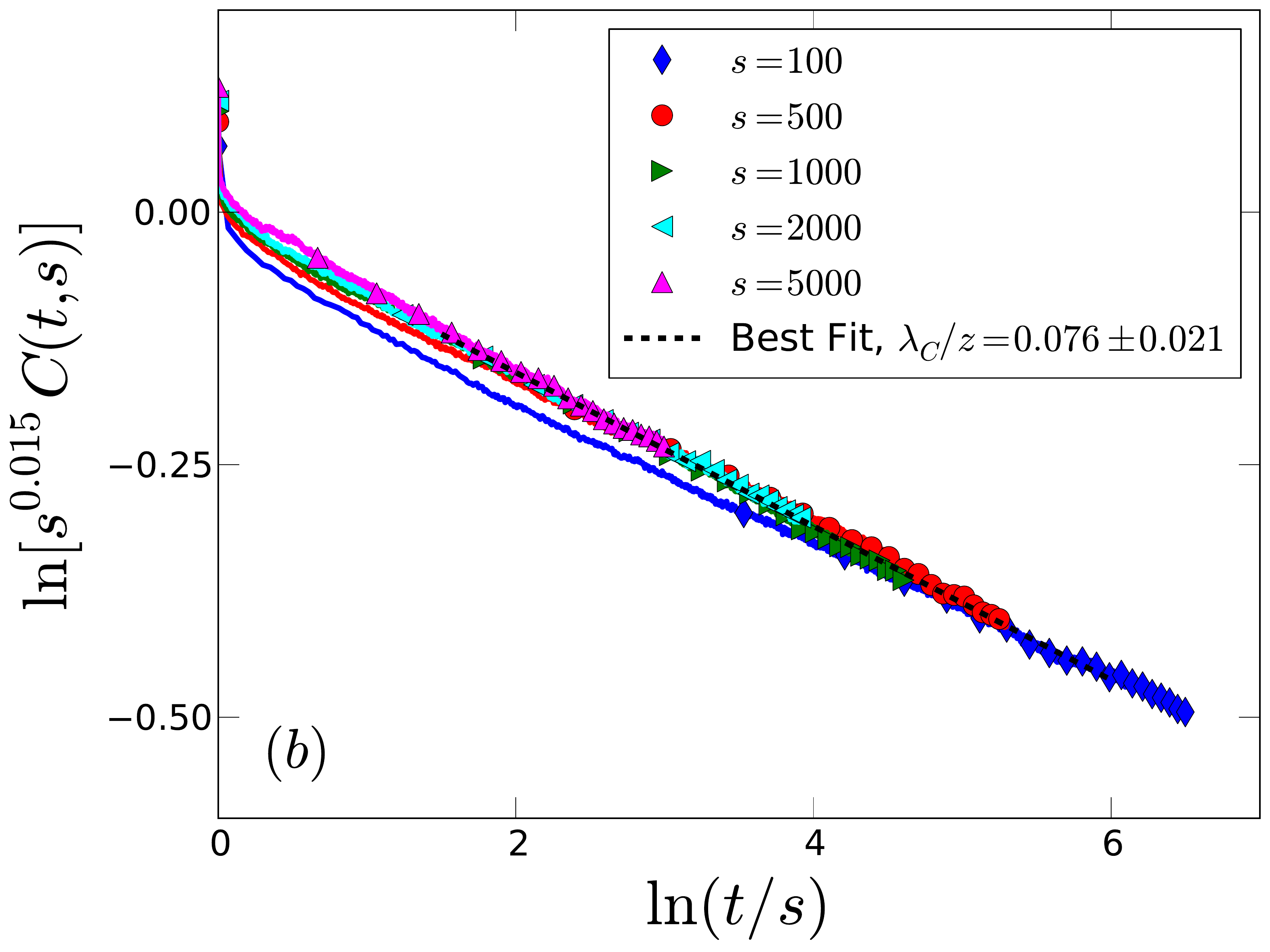} \label{fig:upquench-coulomb-c-aging}}
    \caption{(a) Relaxation and (b) scaling of the two-time carrier density autocorrelation function in the two-dimensional Coulomb glass model after the filling fraction is suddenly increased from $K=0.5$ to $K_{f}=0.54$ (data averaged over 1000 realizations).}
\label{fig:upquench-coulomb}
\end{figure}

For both cases when (a) the charge carrier density abruptly increases from $K=0.5$ to $K_{f}=0.54$ and (b) the density abruptly decreases from $K=0.5$ to $K_{f}=0.46$, the dynamics is again highly dependent on the waiting times chosen as can be seen in Fig.~\ref{fig:upquench-coulomb-c-linear}. Furthermore, the case of density down-quench shows a similar relaxation dynamics to that of the up-quench case in Fig.~\ref{fig:upquench-coulomb-c-linear}.

Dynamical scaling is not present in the relaxed system with a fixed carrier density, while a similar system with random initial conditions displays dynamical scaling with the scaling exponents $b=0.009$ and $\lambda_{C}/z=0.085$. Investigating the dynamical scaling property after the influence of sudden changes in the density of charge carriers reveals that aging scaling is again a property of the system. Due to the similar relaxational dynamics in the cases of density up-quench and down-quench, we confirm that the dynamical aging scaling exponents $b$ and $\lambda_{C}/z$ are equal in both cases. This dynamical scaling behavior is displayed in Fig.~\ref{fig:upquench-coulomb-c-aging}, from which we obtain the values $b=0.015\pm0.003$ and $\lambda_{C}/z=0.076\pm0.021$.

\subsubsection{Bose Glass Model}
\label{sec:aging-Bose-analysis-2}
The relaxational dynamics of the Bose glass model in two dimensions was also analyzed with a similar goal of studying the effects of sudden changes in the density of magnetic flux lines. 

\begin{figure}[h]
  \centering
  \includegraphics[width=0.97\columnwidth]{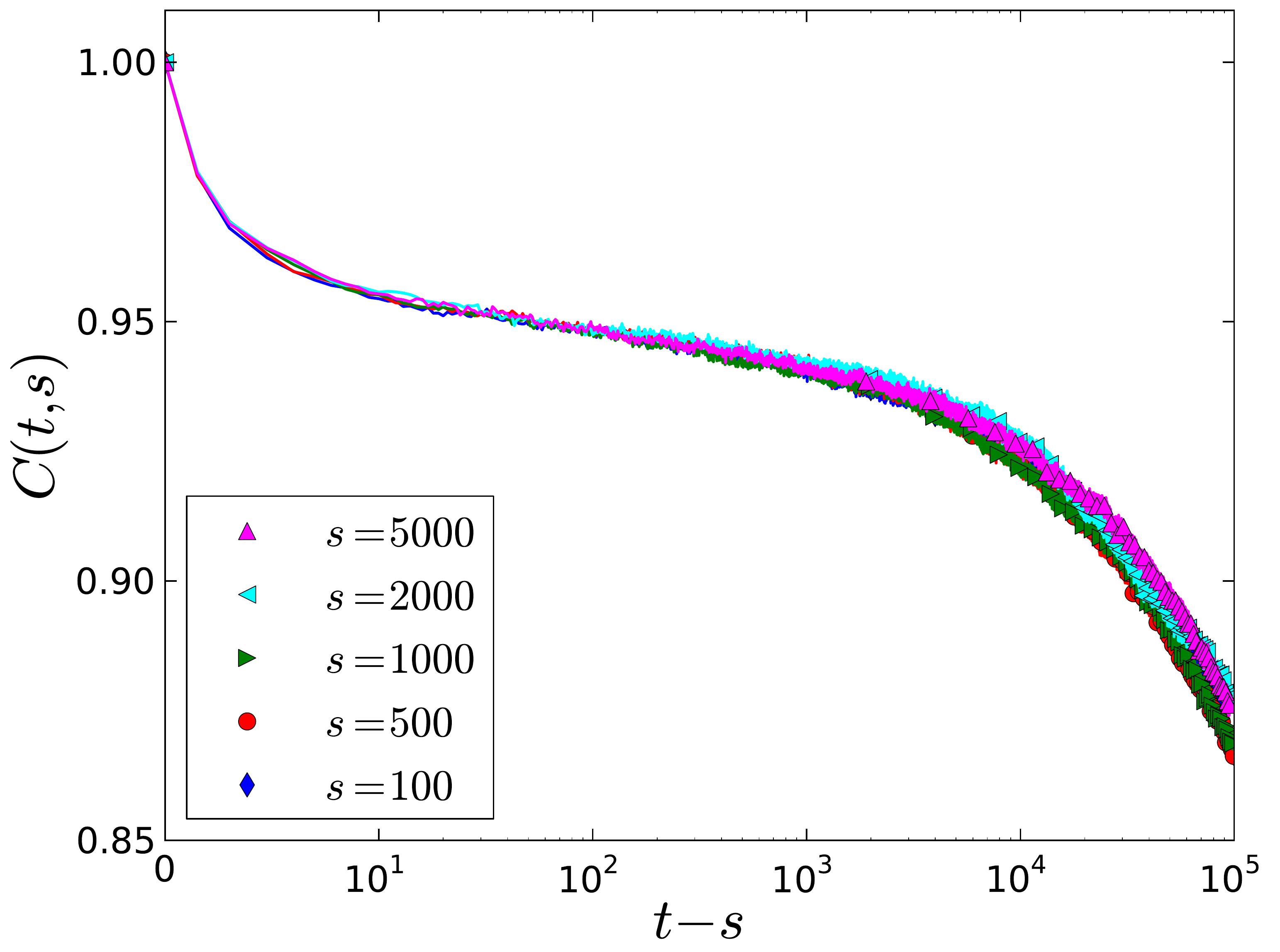}
  \caption{Relaxation of the two-time density autocorrelation function in the two-dimensional Bose glass model with a fixed flux line density $K=0.5$ (data averaged over 2000 realizations).}
  \label{fig:noquench-log-c-glassy}
\end{figure}

\begin{figure}[h]
  \centering
  \includegraphics[width=0.97\columnwidth]{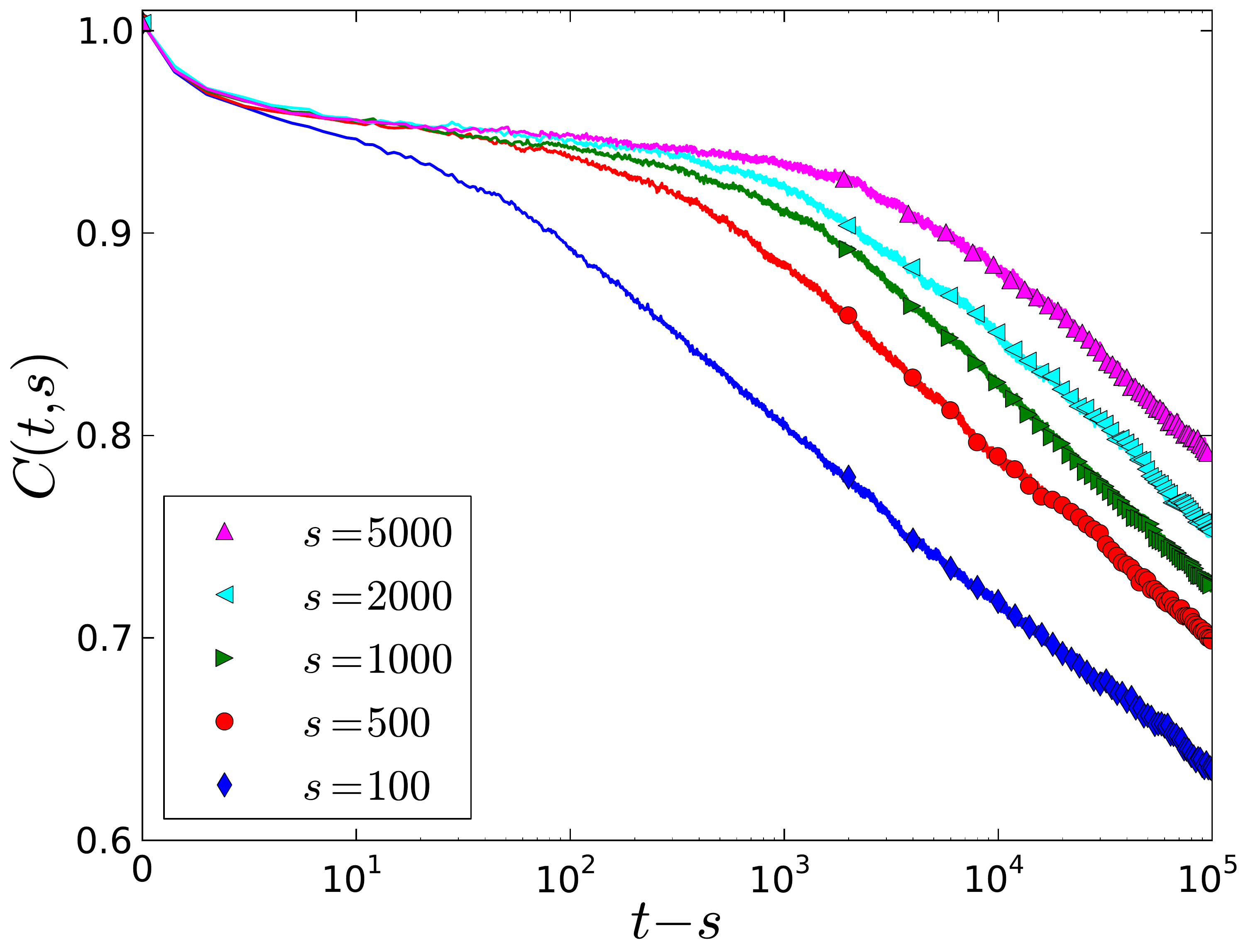}
  \caption{Relaxation of the two-time density autocorrelation function in the two-dimensional Bose glass model after a sudden decrease in the filling fraction from $K=0.5$ to $K_{f}=0.46$. The case when the filling fraction instantaneously increases from $K=0.5$ to $K_{f}=0.54$ displays similar dynamics (data averaged over 1000 realizations).}
  \label{fig:downquench-log-c-glassy}
\end{figure}

When the density of flux lines is kept fixed at $K=0.5$ and the system is relaxed for a sufficiently long initial time $t_{rel}=5\cdot10^{4}$ MCS, the relaxation of the autocorrelation function is very weakly dependent on the waiting times chosen, see Fig.~\ref{fig:noquench-log-c-glassy}.
On the other hand, when the flux line density is suddenly increased from $K=0.5$ to $K_{f}=0.54$ or decreased from $K=0.5$ to $K_{f}=0.46$, we confirm that the system is again dependent on the waiting times $s$. Moreover, both systems similarly display an initial fast relaxation regime and a slower later regime, see Fig.~\ref{fig:downquench-log-c-glassy}.

Dynamical scaling is not present in the relaxed system with a fixed flux line density, while a similar system with random initial conditions displays dynamical scaling with the scaling exponents $b=0.001$ and $\lambda_{C}/z=0.057$. In situations with abrupt increases or decreases in the magnetic flux line density, the following values for these scaling exponents are found: $b=0.005\pm0.001$ and $\lambda_{C}/z=0.057\pm0.013$. 

\begin{figure}[h]
  \centering
  \includegraphics[width=0.97\columnwidth]{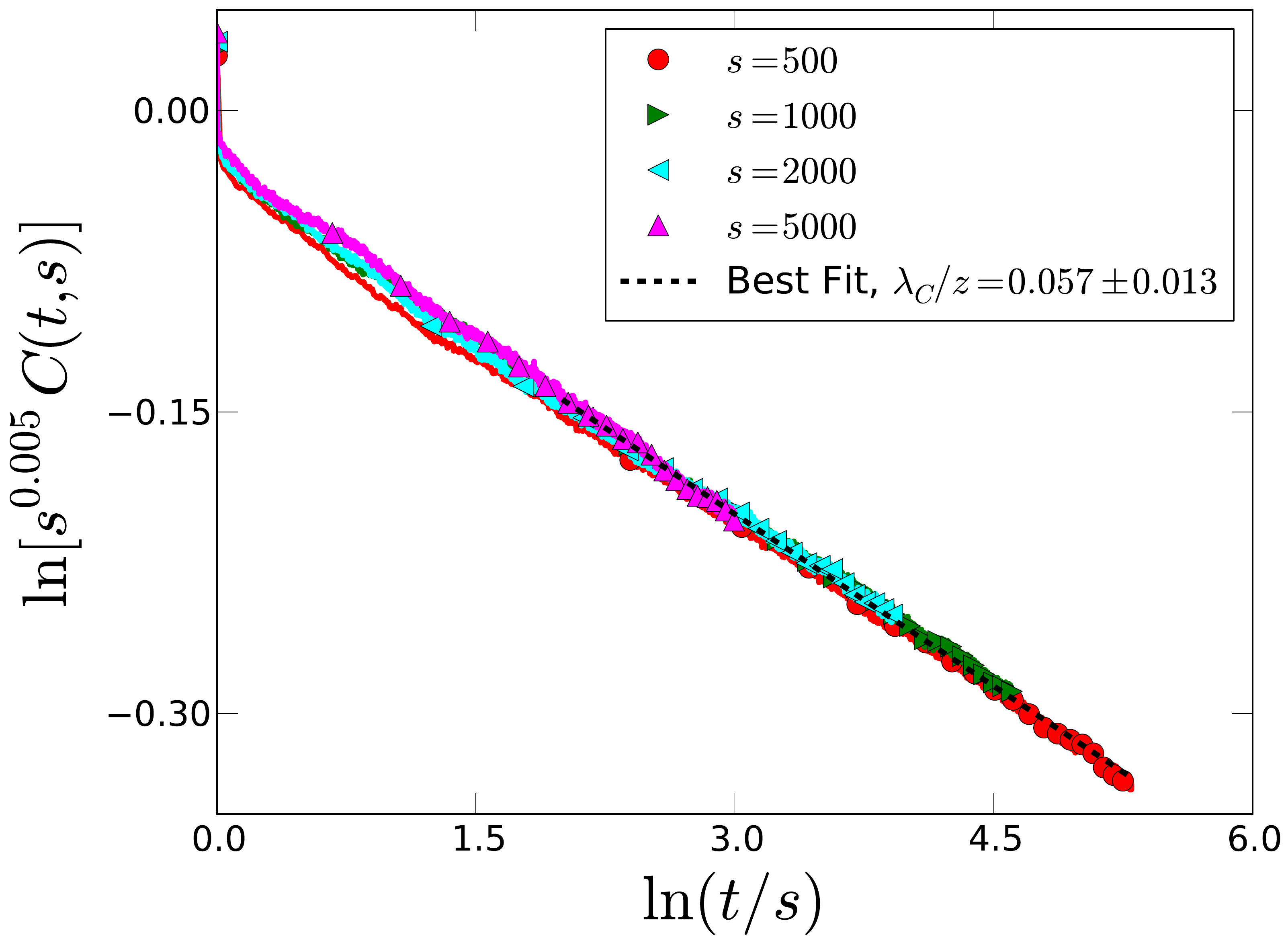}
  \caption{Scaling of the two-time density autocorrelation function in the two-dimensional Bose glass model after the filling fraction is suddenly increased from $K=0.5$ to $K_{f}=0.54$. The case when the filling fraction instantaneously decreases from $K=0.5$ to $K_{f}=0.46$ displays similar dynamics (data averaged over 1000 realizations).}
  \label{fig:upquench-log-c-aging}
\end{figure}

The equal aging scaling exponents $b$ and $\lambda_{C}/z$ in the Bose glass systems when the magnetic flux line density is abruptly increased or decreased implies that this system undergoes similar relaxational dynamics when new flux lines are added to the sample or after some flux lines are removed from the sample due to particle-hole symmetry. This observation was also made in the Coulomb glass model with sudden changes in the charge carrier density.

\section{Conclusions}
\label{sec:con}
We have employed the Coulomb glass model in disordered semiconductors and adapted it to the Bose glass model in type-II superconductors in the presence of extended linear defects to investigate the density of states and the non-equilibrium relaxation properties in the two-dimensional Coulomb and Bose glass systems via Monte Carlo simulations. 

Earlier investigations focused on these systems with zero on-site energies~\cite{Grempel2004,Kolton2005,Shimer2010,Shimer2014}, thus one goal of this study is to analyze the effects of adding random on-site energies from different distributions into the Coulomb and Bose glass systems with a fixed charge carrier/flux line density. We conclude that adding on-site energies from a Gaussian distribution of zero mean and of various widths does not noticeably affect the speed of formation of the Coulomb gap or the effective gap exponent in both models. On the other hand, a flat distribution of a similar width to the one utilized for the Gaussian distribution causes the suppression of the density of states at the chemical potential to become slower in the Coulomb glass model (keeping in mind the possibility that the flat on-site energy distribution might have caused the density of states to freeze out in our small-sized Coulomb glass system in the accessible simulation time scales), while not affecting this feature in the Bose glass model. Furthermore, adding random on-site energies from this flat distribution causes the gap exponent's effective value to approach the mean-field predictions in both models. Such broad on-site energy distributions of widths on the same scale as the correlation-induced Coulomb gap width affect the density of states near the chemical potential in these two models. 

The non-equilibrium relaxation dynamics shows the effects of non-zero random on-site energies as well: The Gaussian-distributed on-site energies do not change the aging scaling exponents, while a wide flat distribution has a pronounced effect on these exponents in both systems. Including random on-site energies from a flat distribution into the Coulomb and Bose glass models causes their relaxation to become faster. This change in the aging scaling exponents implies that these exponents are not universal and in fact depend on various microscopic system parameters~\cite{Shimer2010,Shimer2014}.

Another goal of this study is to analyze the effects of abrupt changes in the density of charge carriers in the Coulomb glass model or magnetic flux lines in the Bose glass model to gauge the systems' response to sudden perturbations. In the cases of density up-quench and down-quench in both models, the system first responds to the quench by shifting the whole DOS curve to higher (up-quench) or lower (down-quench) energies and displaying an asymmetry in the density of states around its minimum value. When the modified system reaches equilibrium, the larger peak shifts to the opposite side of the new chemical potential due to the relaxation of the collective system with the newly-added carriers or the relaxation of the same system after removing some carriers. Since the Coulomb gap exponent is an equilibrium property, one should of course obtain the same exponent in a system with a specific filling fraction regardless of the utilized initial conditions. 

We analyzed the effects of density quenches on the non-equilibrium relaxation dynamics and the aging properties in the Coulomb and Bose glass models. Systems with a fixed density relax for a long time until time translation invariance is displayed by the carrier density autocorrelation function. We observe that dynamical scaling is not a property of finite Coulomb and Bose glass systems at very long times when the density is fixed. On the other hand, when the density is suddenly raised or lowered, the system displays dynamical scaling and aging scaling exponents that are equal in the cases of density up-quench and down-quench in both models. The equal aging scaling exponents when the density of charge carriers or flux lines abruptly changes confirms that the corresponding systems undergo similar relaxational dynamics as a response to the addition of new carriers or the removal of an equivalent number of carriers due to particle-hole symmetry. Within error bars, the aging scaling exponents in the Coulomb and Bose glass models with random initial conditions and after sudden changes in the carrier density coincide, suggesting that structural rearrangements of charge carriers/flux lines are governed by the low-energy equilibrium site energy distributions inside the Coulomb gap. 

\section*{Acknowledgments}
We gladly acknowledge stimulating and helpful discussions with Boris Shklovskii, Matt Shimer, and Shadi Esmaeili. This work was supported by the U.S. Department of Energy, Office of Basic Energy Sciences, Division of Materials Sciences and Engineering, under Grant No. DE-FG02-09ER46613. 

\section*{Authors Contribution Statement}
Hiba Assi carried out the numerical work, analyzed the data, and wrote the first manuscript draft. 
Harshwardhan Chaturvedi assisted in the computational part and coding. 
Michel Pleimling co-advised the research, contributed to data analysis and physical discussions, and assisted in finalizing the paper.
Uwe C. T\"auber proposed the research topic, supervised the research progress and data analysis, and edited the first manuscript draft.

{} 

\end{document}